\documentclass[aip,graphicx,floatfix,preprint]{revtex4-1}

\usepackage{graphicx}
\usepackage{dcolumn}
\usepackage{bm}
\usepackage[utf8]{inputenc}
\usepackage[T1]{fontenc}
\usepackage{mathptmx}
\usepackage{etoolbox}
\usepackage{braket}
\usepackage{mathtools}
\usepackage{dsfont}
\usepackage{xcolor}
\usepackage{amssymb}
\usepackage{textcomp, gensymb}
\DeclarePairedDelimiter\floor{\lfloor}{\rfloor}
\DeclarePairedDelimiter\ceil{\lceil}{\rceil}
\DeclareMathAlphabet{\mathcal}{OMS}{cmsy}{m}{n}
\SetMathAlphabet{\mathcal}{bold}{OMS}{cmsy}{b}{n}

\begin{document}

\title{Electronic Structure Theory with Molecular Point Group Symmetries on Quantum Annealers} 

\author{Joseph Desroches}
\affiliation{ 
Department of Chemistry and Chemical Biology, Northeastern University, Boston, MA 02115, USA}
\affiliation{ 
Department of Physics, Northeastern University, Boston, MA 02115, USA}

\author{Sijia S. Dong*}
\email{s.dong@northeastern.edu}
\affiliation{ 
Department of Chemistry and Chemical Biology, Northeastern University, Boston, MA 02115, USA}
\affiliation{ 
Department of Physics, Northeastern University, Boston, MA 02115, USA}
\affiliation{ 
Department of Chemical Engineering, Northeastern University, Boston, MA 02115, USA}


\begin{abstract}
Quantum computation has the potential to revolutionize quantum chemistry through major speedups to computation times and exponential reduction of computational resources. Here,
we combine the symmetry-adapted Jordan-Wigner encoding based on the full Boolean symmetry group $\mathbb{Z}_2^k$ with our new implementation of the Xia-Bian-Kais (XBK) method for improving the efficiency of electronic structure theory calculations on quantum annealers, particularly by reducing the number of qubits needed to achieve the same accuracy. By providing a more extensive symmetry-adapted encoding (SAE) than previous work, we are able to simulate molecules larger than those previously reported that have been studied using methods developed for quantum annealers and without using an active space. We calculated the potential energy surfaces of H$_2$, LiH, He$_2$, H$_2$O, O$_2$, N$_2$, Li$_2$, F$_2$, CO, BH$_3$, NH$_3$, and CH$_4$, with the largest molecule in the STO-6G basis set requiring 16 qubits with our SAE, and compared them with full configuration interaction results. The application of SAE to the XBK method provides an exponential reduction of the size of the Hilbert space and scales well with the size of the problem. It does not introduce significant additional errors for even or large values of a key variational parameter that determines the number of ancilla qubits used in the XBK method's Hamiltonian embedding, or for certain molecules such as He$_2$ and H$_2$O. We provide an explanation for this behavior and a recommendation on the usage of our method.
\end{abstract}

\pacs{}

\maketitle 

\section{Introduction}

One of the most promising applications of quantum computation is the exact quantum simulation of molecular and material systems. Exact methods, such as exact diagonalization or full configuration interaction (FCI), quickly become intractable for increasingly large many-body systems due to the exponential scaling of both required computational resources (e.g., memory) and time. Quantum computers, on the other hand, are naturally suited to tackle physics simulation problems and do not suffer from the same time and resource scaling issues\cite{feynmanSimulatingPhysicsComputers1982}. Electronic structure theory is a key area that hopes to see major advancement due to the advent of fault-tolerant large-scale quantum computing\cite{mcardleQuantumComputationalChemistry2020}.

Most of the efforts to tackle electronic structure theory problems on quantum computers have been focused on gate-based quantum computers. There are many competing experimental architectures for gate-based quantum computing, with some popular experimental platforms including superconducting, trapped ion, and photonic based approaches. Gate-based quantum computers work through the implementation of quantum "circuits" by the application of quantum "gates" to quantum bits (qubits) in a way analogous to the functioning of modern classical computers \cite{nielsenQuantumComputationQuantum2010}. Efforts to perform quantum chemistry calculations on gate-based quantum computers have revolved around variational quantum algorithms, with the Variational Quantum Eigensolver (VQE) being the most famous example \cite{peruzzoVariationalEigenvalueSolver2014}. VQE and the various methods that have improved upon it (e.g., SSVQE \cite{nakanishiSubspacesearchVariationalQuantum2019}, ADAPT-VQE \cite{grimsleyAdaptiveVariationalAlgorithm2019}) have seen success in performing various quantum chemistry calculations\cite{peruzzoVariationalEigenvalueSolver2014}$^,$\cite{grimsleyAdaptiveVariationalAlgorithm2019}$^,$\cite{nakanishiSubspacesearchVariationalQuantum2019}.

Adiabatic quantum computation is an alternative model for universal quantum computation \cite{aharonovAdiabaticQuantumComputation2005}. Instead of implementing quantum gates in circuits like gate-based quantum computation, adiabatic quantum computation relies on the adiabatic theorem of quantum mechanics to evolve an easy-to-initialize Hamiltonian to a Hamiltonian where the ground state encodes the solution of the desired problem. To perform a calculation on an adiabatic quantum computer, it is only necessary to translate the problem at hand to a Hamiltonian whose ground state encodes the problem's solution. The closest modern experimental realization of an adiabatic quantum computer are D-Wave System's quantum annealers \cite{kingQuantumCriticalDynamics2023}. Of D-Wave Systems' computers, the 5000 qubit Advantage is the largest, with approximately five times more qubits available than the largest modern gate-based quantum computers \cite{kingQuantumCriticalDynamics2023}. The governing Hamiltonian for D-Wave Systems' quantum annealer devices is 
\begin{equation} \label{eq:1}
    \hat{H}(s) = - \frac{A(s)}{2} \bigg( \sum_{i} \hat{\sigma}^i_x \bigg) + \frac{B(s)}{2} \bigg( \sum_{i} h_i \hat{\sigma}^i_z + \sum_{ij} J_{ij} \hat{\sigma}^i_z \hat{\sigma}^j_z \bigg)
\end{equation}

where $\sigma_{x}^i$, $\sigma_y^i$, and $\sigma_z^i$ are the Pauli matrices acting on the $i$-th qubit, $A(s)$ and $B(s)$ are the annealing schedule such that $A(s=0)=1, \: B(s=0)=0$ and $A(s=1)=0, \; B(s=1)=1$. The initial Hamiltonian (the multiplier of $-A(s)/2$ in Eq. (1)) is experimentally simple to prepare, with a ground state of $\ket{\uparrow,\uparrow,\dots,\uparrow}$. The final Hamiltonian (the multiplier of $B(s)/2$ in Eq. (1)) is the problem Hamiltonian, in which one encodes the problem of interest into the ground state. For D-Wave Systems' quantum annealers, the problem Hamiltonian can only take an Ising form, or equivalently, a quadratic unconstrained binary optimization (QUBO) form. Hence, a quantum annealer can only handle stoquastic Hamiltonians while a true (universal) adiabatic quantum computer could handle an arbitrary non-stoquastic Hamiltonian.

Several efforts have been devoted to developing algorithms for electronic structure theory calculations on quantum annealers by transforming the electronic structure problem into a suitable form. The most direct translation is the Xia-Bian-Kais (XBK) method, which encodes the molecular Hamiltonian into the physical embeddings of a quantum annealer\cite{xiaElectronicStructureCalculations2018}. One can also use the Qubit Coupled Cluster (QCC) method to map the electronic structure problem to an optimization problem, of which quantum annealers are a natural application\cite{geninQuantumChemistryQuantum2019}. Previous work has also used the Quantum Annealing Eigensolver (QAE), a replacement of a standard classical eigensolver, to address electronic structure, vibrational, lattice gauge theory, and scattering problems on quantum annealers \cite{teplukhinElectronicStructureDirect2020}$^,$\cite{teplukhinCalculationMolecularVibrational2019}$^,$\cite{teplukhinComputingMolecularExcited2021}$^,$\cite{teplukhinSolvingComplexEigenvalue2020}. 

On an alternative front, several efforts have been made to reduce the complexity of the electronic structure problem on quantum computers through the use of symmetry-adapted encodings (SAE) \cite{bravyiTaperingQubitsSimulate2017}$^,$\cite{setiaReducingQubitRequirements2020}$^,$\cite{picozziSymmetryadaptedEncodingsQubit2023}. These SAE can reduce the number of qubits needed to simulate Hamiltonians by finding redundant terms caused by underlying physical symmetries of the systems. Previous work has found success in reducing qubit requirements for electronic Hamiltonians by investigating $\mathbb{Z}_2$ symmetries \cite{bravyiTaperingQubitsSimulate2017} and molecular point group symmetries \cite{setiaReducingQubitRequirements2020}$^,$\cite{picozziSymmetryadaptedEncodingsQubit2023}. 

In this work, we combine the symmetry-adapted Jordan-Wigner (JW) encoding based on the full Boolean symmetry group $\mathbb{Z}_2^k$ with the XBK method to calculate the potential energy surfaces of a set of small molecules using diagonalization. Because our study uses diagonalization to find the ground state of our XBK-transformed Ising Hamiltonians, we are able to avoid hardware related errors associated with modern day quantum annealers and instead to evaluate rigorously the numerically exact performance of our symmetry-adapted XBK method. Our work is the first to extend the consideration of symmetries to the full Boolean symmetry group based on the molecular point groups. Previous work has calculated ground state properties for a small list of molecules (H$_2$, H$_3^+$, LiH, He$_2$, HeH$^+$, and H$_2$O) on quantum annealers with a varyingly effective symmetry reductions \cite{streifSolvingQuantumChemistry2019}$^,$\cite{copenhaverUsingQuantumAnnealers2021}. However, since we employ a more extensive symmetry-adapted encoding than in previous works, we are able to calculate potential energy surfaces for molecules larger than those previously reported.\cite{streifSolvingQuantumChemistry2019}$^,$\cite{copenhaverUsingQuantumAnnealers2021} Furthermore, we extensively demonstrate and provide an explanation for uncharacteristic behaviors of the XBK method when SAE is applied. A similar behavior for H$_2$ and LiH was previously seen, but no explanation was given\cite{streifSolvingQuantumChemistry2019}.

\section{Methods}
\subsection{Electronic Structure Theory on Quantum Computers}

Electronic structure theory is primarily concerned with finding the eigenstates and eigenenergies of the electronic Hamiltonian. In general, this is an interacting quantum many-body problem, which is exponentially hard to solve on a classical computer for an increasing number of particles due to the size of the Hilbert space scaling as $2^N$ for $N$ electrons.

One formalism that is particularly well-suited to solving interacting many-body problems is the second quantization formalism of quantum mechanics. In this formalism, the electronic Hamiltonian is formulated in terms of the fermionic creation and annihilation operators $\hat{a}^\dagger$ and $\hat{a}$, which populate or depopulate certain electronic orbitals and obey the standard canonical anticommutation relations. Typically, to construct the second quantization electronic Hamiltonian, one chooses a particular basis set and calculates the one- and two-body electron integrals $h_{ij}$ and $h_{ijkl}$ and writes
\begin{equation} \label{eq:2}
    \hat{H} = \sum_{ij} h_{ij} \hat{a}_i^\dagger \hat{a}_j + \sum_{ijkl} h_{ijkl} \hat{a}_i^\dagger \hat{a}_j^\dagger \hat{a}_k \hat{a}_l
\end{equation}

In the second quantization formalism, within a basis set with $M$ total spin-orbitals, an arbitrary state of the electronic Hamiltonian can be expressed in the form $\ket{\psi} = \ket{n_1,n_2,\dots,n_j,\dots,n_M}$, where each $n_j \in \{0,1\}$ represents the occupation of the $j$-th spin-orbital. 

In order to translate Eq. \ref{eq:2} into a form suitable for quantum computation, it is necessary to map the fermionic creation and annihilation operators to qubit operators. Many such so-called qubit transformations can map Eq. \ref{eq:2} in a suitable form, including, most notably, the Jordan-Wigner transformation, the Bravyi-Kitaev transformation, and the parity transformation \cite{seeleyBravyiKitaevTransformationQuantum2012}$^,$\cite{tranterBravyiKitaevTransformation2015}$^,$\cite{bravyiFermionicQuantumComputation2002}. For the interested reader, a comprehensive discussion of these transformations can be found in  \cite{bravyiTaperingQubitsSimulate2017}. Furthermore, it is possible to include information about the qubit Hamiltonian's internal symmetries in the fermion-to-spin mapping. We detail the efficient algorithm we use to apply SAE based on the full Boolean symmetry group, developed in \cite{picozziSymmetryadaptedEncodingsQubit2023}, in Sec. \textbf{IIB}.

Once a qubit transformation has been applied to Eq. \ref{eq:2}, the electronic Hamiltonian takes the form
\begin{equation} \label{eq:3}
    \hat{H} = \sum_{i} \sum_{\alpha} h^i_\alpha \hat{\sigma}^i_\alpha + \sum_{ij} \sum_{\alpha \beta} h^{ij}_{\alpha \beta} \hat{\sigma}^i_\alpha \hat{\sigma}_\beta^j + \dots
\end{equation}

where $\hat{\sigma}^i_\alpha$, $i \in \{1,2,\dots,m\}$, $\alpha \in \{x, y, z\}$ are the Pauli matrices that act on the $i-th$ qubit. Rewriting Eq. \ref{eq:3} in terms of the so-called Pauli words 
\begin{equation} \label{eq:4}
    \hat{H} = \sum_{i} \eta_i P_i
\end{equation}

where each Pauli word is an element of the Pauli group $P_i \in \mathcal{P}_m = \pm \{\mathds{1}, \hat{\sigma}_x, \hat{\sigma}_y, \hat{\sigma}_z\}^{\otimes m} $. Note that, as $m \leq M$, the number of qubits does not need to match the number of spin orbitals in the chosen basis. Once the Hamiltonian has been transformed to a qubit form, $\hat{H}$ acts on $m$-qubit basis states of the form
\begin{equation} \label{eq:5}
    \ket{\phi} = \prod_{i=1}^m \ket{z_i} = \ket{z_1, z_2, \dots, z_m}
\end{equation}

where each $z_i = 0$ represents a spin-up qubit and $z_i = 1$ represents a spin-down qubit. Any $m$-qubit state can then be written as a linear combination of the $2^m$ basis states as $\ket{\Psi} = \sum_{i}^{2^m} a_i \ket{\phi_i}$.

To perform the electronic structure theory on quantum annealers it is necessary to further process Eq. \ref{eq:4}. As quantum annealers are restricted to handling Hamiltonians in the form of Eq. \ref{eq:1}, it is necessary to map Eq. \ref{eq:6} to an Ising-type Hamiltonian. One such popular method to accomplish this is the XBK transformation\cite{xiaElectronicStructureCalculations2018}. To the best knowledge of the authors, the XBK method is the only transformation that directly maps Eq. \ref{eq:4} to an Ising-type Hamiltonian. The full details of the XBK transformation can be found in  \cite{xiaElectronicStructureCalculations2018}, and a summary of the method can be found in section \textbf{IV}.

\subsection{Reducing Qubit Requirements with Molecular Point Group Symmetries}

Using physical symmetries, such as parity \cite{bravyiTaperingQubitsSimulate2017}, number conservation \cite{bravyiTaperingQubitsSimulate2017}$^,$\cite{setiaReducingQubitRequirements2020}$^,$\cite{picozziSymmetryadaptedEncodingsQubit2023}, and molecular point group symmetries \cite{setiaReducingQubitRequirements2020}$^,$\cite{picozziSymmetryadaptedEncodingsQubit2023}, it is possible to reduce the number of qubits needed to simulate a Hamiltonian. In \cite{bravyiTaperingQubitsSimulate2017}, an algorithm was presented that efficiently finds the $\mathbb{Z}_2$ symmetries of a Hamiltonian and then tapers off qubits depending on how many $\mathbb{Z}_2$ symmetries the Hamiltonian has. In \cite{setiaReducingQubitRequirements2020}, a similar method was extended to include the molecular point group symmetries and provided physical justification for the $\mathbb{Z}_2$ symmetries found with the algorithm in \cite{bravyiFermionicQuantumComputation2002}. However, no efficient algorithm to automate this extended symmetry-adapted encoding for an arbitrary molecular system was presented. Later, in \cite{picozziSymmetryadaptedEncodingsQubit2023}, an efficient algorithm was introduced that automates this procedure, although in an alternative formulation. The method developed in \cite{picozziSymmetryadaptedEncodingsQubit2023} makes direct use of knowledge of the character table of the Boolean point group, a subgroup of the molecular point group, and knowledge of the representations of the molecular orbitals \cite{picozziSymmetryadaptedEncodingsQubit2023}. Compared to previous methods for applying SAE, this algorithm runs at minimal computational cost \cite{picozziSymmetryadaptedEncodingsQubit2023}.

In the algorithm proposed in \cite{picozziSymmetryadaptedEncodingsQubit2023}, which we utilize in this study, the Boolean point group symmetries on the Jordan-Wigner basis have a simple form given that the associated qubits signify symmetry-adapted molecular spin orbitals \cite{picozziSymmetryadaptedEncodingsQubit2023}. To build them, one must examine the column for the particular point group element in the character table for the relevant symmetry group. A specific symmetry operator will act as a $\sigma^z$ on the $j$-th qubit if the associated spin-orbital is in a representation that is antisymmetric with respect to the symmetry (i.e., $-1$ in the character table) and will act trivially on the $j$-th qubit if the associated spin-orbital is in a representation that is symmetric with respect to the symmetry (i.e., $+1$ in the character table). The resulting operator has its positive eigenspace spanned by computational basis states corresponding to Slater determinants that are symmetric under the symmetry and has its negative eigenspace spanned by computational basis states corresponding to Slater determinants that are antisymmetric under the symmetry. Using these guidelines, one can promptly construct the qubit representations of the Boolean point group symmetries solely from the point group character table. 

With the Jordan-Wigner transformation, the Slater determinants are uniquely matched with the computational basis states \cite{picozziSymmetryadaptedEncodingsQubit2023}. Spin orbitals exhibit either symmetric or antisymmetric behavior relative to a Boolean symmetry operator, thus allowing Slater determinants to assume one of these two forms. Consequently, a Boolean symmetry segregates the entire set of possible Slater determinants into two groups: one group in the symmetric eigenspace and the other in the antisymmetric eigenspace. For the Boolean symmetry group $\mathbb{Z}_2^k$, where $k$ represents the number of independent generators, this results in $k$ linear equation constraints that determine the occupancies of spin orbitals.

Following \cite{picozziSymmetryadaptedEncodingsQubit2023}, let $g_0, \dots, g_{k-1}$ be $k$ independent generators of the Boolean symmetry group $\mathbb{Z}_2^k$, let $\mathcal{A}_0, \dots, \mathcal{A}_{k-1}$ be sets of spin-orbitals that are antisymmetric with respect to the symmetry of $\mathbb{Z}_2^k$, i.e., $g_0, \dots, g_{k-1}$. Furthermore, let $c_0, \dots, c_{k-1} \in \{ 0,1 \}$ be zero if the target irreducible is symmetric to the symmetry and 1 if it is anti-symmetric (i.e., $\pm 1$ in the character table), and let $p_0, \dots, p_{k-1} \in \{0,1\}$ be the spin orbital occupancies/ the qubit states in the Jordan-Wigner basis. This leads to a set of $k$ binary constraints 
\begin{equation} \label{eq:6}
    \bigoplus_{p_j\in\mathcal{A}_i} p_j  = c_i
\end{equation}

where $\oplus$ represents addition modolo 2. This implies that the occupancies of the $k$ spin orbitals in $\bigcup_{i=1}^{k} \mathcal{A}_i$ are redundant. Therefore, if an electronic Hamiltonian is invariant under a Boolean symmetry $\mathbb{Z}_2^k$, one is able to remove $k$ spin orbitals (equivalently, basis states in the Jordan-Wigner encoding) and reduce the Hilbert space by $k$ qubits. 

\subsection{The Xia-Bian-Kais (XBK) Method}

The XBK method uses ancilla qubits to simulate non-Ising-type interactions in terms of Ising interactions allowed in Eq. \ref{eq:1}. The XBK transformation maps the $m$-qubit problem of a non-stoquasitc Hamiltonian $\hat{H}$ to a $rm$-qubit problem of a stoquastic Hamiltonian $\hat{H}'$. $r$ plays the role of a variational parameter, in which $r$ can be increased to increase the Hilbert space of $\hat{H}'$ and account for a larger number of non-Ising-type interactions that may be present in $\hat{H}$. In the limit of large $r$, the XBK method becomes exact, with all non-Ising type interactions accounted for in the (exponentially) larger Hilbert space of the $rm$-qubit Hamiltonian. Therefore, by increasing $r$ one can achieve arbitrary accuracy in determining the eigenspectrum of the original Hamiltonian $\hat{H}$ using only Ising-type interactions present in $\hat{H}'$. 

There are two steps to accomplish this mapping \cite{xiaElectronicStructureCalculations2018}. The first step is to map the wavefunction $\ket{\Psi}=\sum_{i} a_i \ket{\phi_i}$ acting on $m$ qubits to a wavefunction $\ket{\Phi}$ acting on $rm$ qubits. This is accomplished by repeating the basis $\{\ket{\phi_i}\}$ $b_i$ times, where $b_i$ is determined by the coefficients $a_i$ through the relationship
\begin{equation} \label{eq:7}
    a_i \approx \frac{b_i S(b_i)}{\sqrt{\sum_{m}b_m^2}}
\end{equation}

where $S(b_i)$ is a sign function and $\sum_i b_i = r$. The sign function $S(b_i)$ is required as the coefficients $a_i$ can be positive or negative, but the repetition times of a basis (i.e., $b_i$) must be positive \cite{xiaElectronicStructureCalculations2018}. After this mapping, the any $rm$-qubit state $\ket{\Phi}$ can be represented as 
\begin{equation} \label{eq:8}
    \ket{\Phi} = \otimes_i \otimes_{j=1}^{b_i} \ket{\phi_i}
\end{equation}

The second part of the mapping is to map $\hat{H}$ to $\hat{H}'$. As the Ising Hamiltonian, i.e., Eq \ref{eq:1}, either incorporates only a single $\hat{\sigma}^z$ operator or a product of two $\hat{\sigma^z}$ operators, all terms that do not match this condition must be mapped. The mapping used in the XBK method that accomplishes this transforms the operators   
\begin{equation} \label{eq:9}
    \begin{aligned}
        \hat{\sigma}_x^i &\rightarrow \frac{1 - \hat{\sigma}_z^{i_j} \hat{\sigma}_z^{i_k}}{2}, & \quad \hat{\sigma}_y^i &\rightarrow \frac{\hat{\sigma}_z^{i_k} - \hat{\sigma}_z^{i_j}}{2}, \\ 
        \hat{\sigma}_z^i &\rightarrow \frac{\hat{\sigma}_z^{i_j} + \hat{\sigma}_z^{i_k}}{2}, & \quad I_i &\rightarrow \frac{1 + \hat{\sigma}_z^{i_j} \hat{\sigma}_z^{i_k}}{2}
    \end{aligned}
\end{equation}

The notation $\hat{\sigma}_z^{i_j}$ represents the $\hat{\sigma}_z$ operator acting on the $i$-ith qubit in the $j$-th $m$-qubit subspace (i.e., the $[(j-1)n+i]$-th qubit out of the $rm$ total qubits). Applying Eq. (11) to every operator in Eq. (7) will yield a "sub-Hamiltonian" $\hat{H}^{(i,i)}$ that acts on $2m$ qubits. To account for the possible sign permutations, one introduces a parameter $0\leq p \leq \floor{\frac{r}{2}}$ such that\cite{xiaElectronicStructureCalculations2018}$^,$\cite{copenhaverUsingQuantumAnnealers2021} 
\begin{equation} \label{eq:10}
    S_p(i) = \begin{cases} 
    -1, & \text{if } i \leq p \\
    1, & \text{else}
    \end{cases}
\end{equation}

To construct the $\ceil{\frac{r}{2}}$ possible $rm$-qubit Hamiltonians, one sums over the sub-Hamiltonians $\hat{H}^{(i,j)}$ for $0\leq i,j \leq r$ for each possible $p$ value as
\begin{equation} \label{eq:11}
    \hat{H}_{p}^{'} = \sum_{i,j}^r \hat{H}^{(i,j)} S_p(i) S_p(j)
\end{equation}

Each $\hat{H}_{p}^{'}$ explores a sector of the full $rm$-qubit eigenspace. In order to find the ground state of the complete Hamiltonian $\hat{H}$, it will be necessary to iterate over $p$ to find which sector contains the lowest energy.

As shown in Xia et al. \cite{xiaElectronicStructureCalculations2018}, if $\hat{H}$ has an eigenvalue of $\lambda$, then $\hat{H}_{p}^{'}$ will have an eigenvalue $\lambda' = \lambda \sum_m b_m^2$. Thus, it is important to construct an operator $\hat{C}_p$ that counts $\sum_m b_m^2$ for each $p$ value, 
\begin{equation} \label{eq:12}
    \hat{C}_p = \sum_{\pm} \left[ \sum_{i=1}^r \left( S_p(i) \prod_{k=1_i}^{m_i} \frac{1 \pm \sigma_z^k}{2} \right) \right]^2
\end{equation}

In practice, $\sum_m b_m^2$ is determined by applying the $\hat{C}_p$ operator to the $rm$-qubit ground state wavefunction $\ket{\Phi^{(0)}_p}$ for each $\hat{H}_p^{'}$. Hence, the procedure to find the ground state of $\hat{H}$ using the XBK method involves iterating over all possible values of $p$. For each $p$, one must calculate the ground state wavefunction $\ket{\Phi_p^{(0)}}$ and energy $E_{p}^{(0)}$, construct and apply the operator $\hat{C}_p \ket{\Phi_p^{(0)}} = \sum_m b_m^2$, and calculate the corresponding $\hat{H}$ eigenvalue $E_{p} = E_p^{(0)}/\sum_m b_m^2$. The smallest of the set of $\{E_{p}, \: 0\leq p \leq \floor{\frac{r}{2}}\}$ is then the true ground-state energy of the Hamiltonian $\hat{H}$. Because we explicitly specify the eigenspace of our Ising Hamiltonian by choosing a particular $p$ value, the remaining eigenspecturm does not necessarily contain physical states that correspond to the excited states of the original Hamiltonian. To our knowledge, there is currently no method to determine the excited states of a Hamiltonian using the XBK transformation due to this reason.

There exists an optimization algorithm to perform this procedure to calculate $\ket{\Phi_p^{(0)}}$ and $E_p^{(0)}$ on a quantum annealing device, as developed by Xia et al. and implemented in a GitHub repository by Copenhaver et al. \cite{xiaElectronicStructureCalculations2018}$^,$\cite{copenhaverUsingQuantumAnnealers2021}. The full details of this algorithm can be found in the original work \cite{xiaElectronicStructureCalculations2018}, and with an implementation in Python \cite{copenhaverUsingQuantumAnnealers2021}. However, in our work, we develop a different implementation of the XBK method. Our implementation is equivalent to the optimization algorithm by Xia et al., reproducing its exact same behavior for all molecules and cases considered, as shown in the supplementary material. In the original work of the XBK method\cite{xiaElectronicStructureCalculations2018}, the ground-state energy and wave function of the $p$-th Ising Hamiltonian is obtained through an iterative procedure that continues until $(\hat{H}^{'}_{p} - \lambda \hat{C}) < 0$, where $\lambda$ evolves with each iterative step. In our implementation, we directly diagonalize $\hat{H}^{'}_{p}$ to obtain the exact energy and wavefunction of the ground state. We achieve this using the Implicitly Restarted Lanczos Method (IRLM) as implemented in the SciPy Python package. We use this classical method for this for
two main reasons. First, because we avoid embedding our Hamiltonians into the quantum annealer, we avoid hardware-related errors and instead are able to produce numerically exact results. Second, due to the added requirements of quadratization, quantum error correction, and embedding, our Hamiltonians that can be handled classically (that is, through exact diagonalization) require too many qubits to run on quantum annealers currently available \cite{copenhaverUsingQuantumAnnealers2021}. With our implementation, one can handle significantly more molecules and molecular geometries with massively parallel classical computing, as compared to the limited accessible runtime on real quantum annealing devices. As a consequence, we can distill and investigate the behaviors of our symmetry-adapted XBK method theoretically separate from the limitations of the quantum hardware. 

\subsection{The Xia-Bian-Kais Method with a Symmetry-Adapted Jordan-Wigner Encoding} 

Here, we detail the method we use to construct the symmetry-adapted Ising Hamiltonian and calculate the ground-state potential energy surfaces of the molecules presented in this work. We implement our method in a computer code in Python, which is available at \url{https://github.com/sdonglab/SymmetryAdaptedXBK}. Our Python code combines and extends upon previous code repositories developed by \cite{copenhaverUsingQuantumAnnealers2021} to perform the XBK transformation and \cite{picozziSymmetryadaptedEncodingsQubit2023} to apply SAE based on the full Boolean symmetry group. 

\begin{figure*}[t!]
    \centering
    \includegraphics[width=\textwidth]{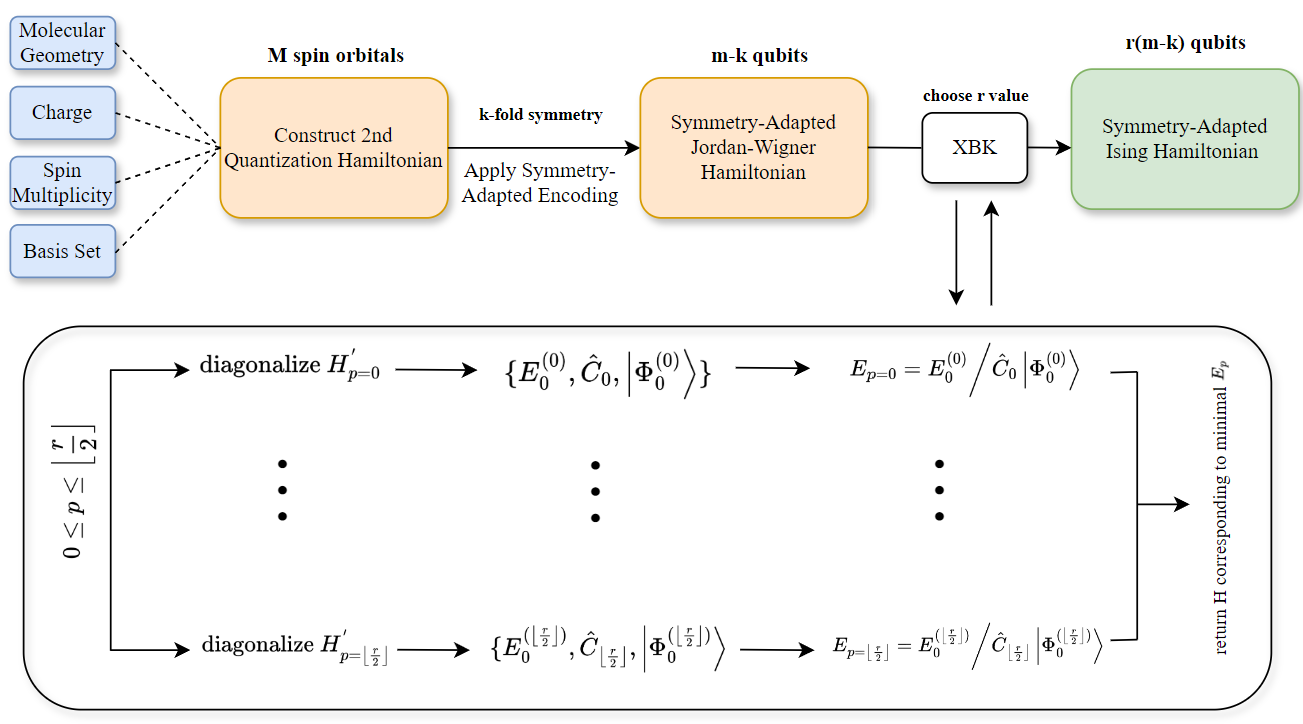}
    \caption{An overview of our algorithm. The input of our algorithm includes the molecular geometry, the molecular charge, the spin multiplicity, the basis set (to construct the one- and two-electron integrals), and the variational parameter $r$. We perform the XBK with our own implementation, which uses direct diagonalization to minimize $E_p$. Once we have obtained the correct Ising Hamiltonian, which contains our ground state, we can either find its ground state or its extended spectrum (the eigenvalues/eigenvectors beyond the ground state). }
    \label{fig_overview}
\end{figure*}

We provide a diagrammatic overview of our algorithm in Fig. \ref{fig_overview}. The main algorithm of our code is summarized as follows: 
\begin{enumerate}
    \item We calculate the one- and two-electron integrals using a classical quantum chemistry package in OpenFermion-PySCF \cite{sunPySCFPythonbasedSimulations2018} for a specific choice of molecular geometry. We then use these to construct the electronic Hamiltonian, i.e., Eq. (3) for our molecular system using OpenFermion \cite{mccleanOpenFermionElectronicStructure2020}.

    \item We use a symmetry adapted Jordan Wigner transformation \cite{picozziSymmetryadaptedEncodingsQubit2023} to write down the symmetry-adapted qubit Hamiltonian $\hat{H}$. This symmetry adapted Jordan Wigner transformation makes use of the molecular system's Boolean molecular point symmetries, as well as the conservation properties including parity, total electron number, etc.

    \item We choose a particular $r$ value for the XBK method. We then iterate $0\leq p \leq \floor{\frac{r}{2}}$. For each $p$, we calculate $\hat{H}_{p}^{'}$ and $\hat{C}_p$. We then use the Implicitly Restarted Lacnzos Method as implemented in the SciPy Python package to find the ground-state energy $E_{p}^{(0)}$ and wavefunction $\ket{\Phi_p^{(0)}}$ of $\hat{H}_{p}^{'}$. We then calculate the corresponding eigenvalue of $\hat{H}$ by applying $\hat{C}_p \ket{\Phi_p^{(0)}} = \sum_m b_m^2$, and dividing $E_{p} = E_p^{(0)}/\sum_m b_m^2$.

    \item Once we have iterated over all $p$, we find the true ground-state energy of $\hat{H}$, $E_0 = \min \{E_p, \: \:0\leq p \leq \floor{\frac{r}{2}} \} = E_{p'}^{(0)}$. We then return the $rm$-qubit ground state wavefunction $\ket{\Phi_{p'}^{(0)}}$ corresponding to the $p$ value that minimizes $E_p$.
\end{enumerate}

When we calculate the ground state of electronic Hamiltonians without SAE to compare the symmetry-adapted Hamiltonians to, we use a standard Jordan-Wigner transformation in place of the symmetry-adapted Jordan-Wigner transformation in the second step of our procedure. We implement the standard JW transformation using the OpenFermion Python package. When we calculate the full eigenspectrum of the Ising Hamiltonians produced with the XBK transform, we employ the LAPACK routines \_syevd, \_heevd as implemented in the NumPy Python package, as compared to the IRLM implementation in SciPy. Finally, all classical calculations were performed on a high-performance computing cluster.

\section{Results}

We applied our algorithm to obtain bond dissociation curves for H$_2$, LiH, He$_2$, H$_2$O, BH$_3$, and NH$_3$. Results for additional molecules can be found in the supplementary material. 

\subsubsection{Molecular Hydrogen - H$_2$}

Molecular hydrogen (H$_2$) is the quintessential test case for electronic structure theory, as its relatively simple structure and small size allow it to be simulated very rapidly. H$_2$ is the smallest neutral molecule, with only two electrons and protons, and allows for exact solutions in FCI within a given basis set providing a crucial benchmark. The potential energy curve of H$_2$ exhibits several important features, including the bonding region, dissociation limit, and avoided crossings between electronic states, making it a comprehensive yet computationally easy test for our method. Furthermore, H$_2$ has previously been simulated with exact diagonalization \cite{xiaElectronicStructureCalculations2018}, simulated, and quantum annealing \cite{streifSolvingQuantumChemistry2019}$^,$\cite{copenhaverUsingQuantumAnnealers2021}, making it an ideal comparison point for comparing our method. Molecular hydrogen is also a high-symmetry system, maintaining a $\mathbb{Z}_2^3$ symmetry due to its $P^\uparrow$, $P^\downarrow$, and $C_{2y}$ symmetries. The $P^{\{\uparrow,\downarrow\}} = (-1)^{\hat{N}}$ parity symmetries for spin-up and spin-down electrons respectively correspond to the exponentiation of the number operators $\hat{N}^{\{\uparrow,\downarrow\}}$, which we assume to be conserved for our work here. The $C_{2y}$ symmetries correspond to the 180$\degree$ 
 rotation symmetry about the $y$-axis.

\begin{figure}[t!]
    \includegraphics[width=\columnwidth] {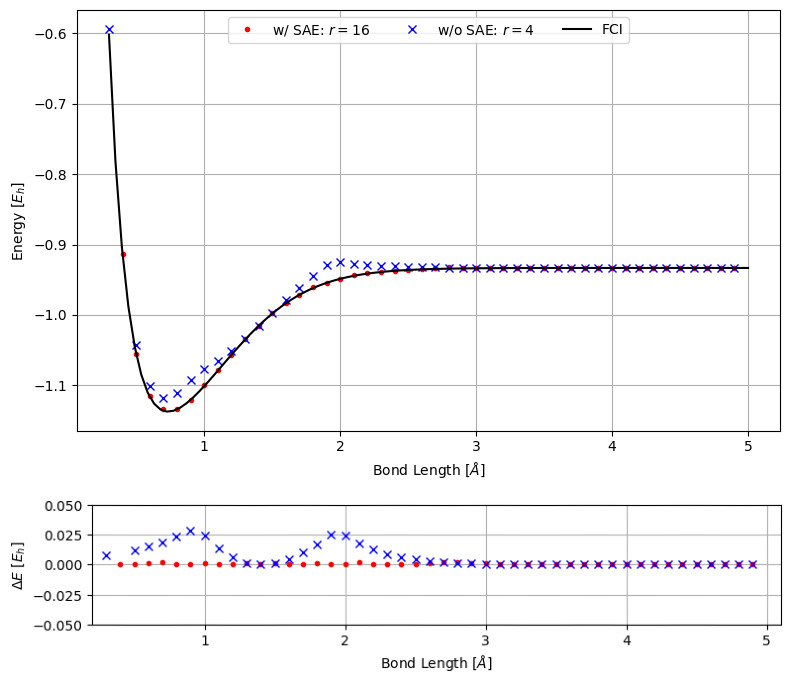}
    \caption{The ground-state bond dissociation energy curve of H$_2$, calculated by the diagonalization of the $r=16$ symmetry-adapted Ising Hamiltonian in the STO-3G basis, compared with the bond dissociation energy curve from the non-symmetry-adapted XBK method that requires the same number of qubits. }
    \label{fig_h2}
\end{figure}

First, we demonstrate that, compare to the non-symmetry-adapted method, our method allows for higher-accuracy calculations of the potential energy surface of molecular hydrogen with the same number of qubits. Molecular hydrogen is an ideal example for this: Because of the small number of qubits needed to simulate both its symmetry-adapted and non-symmetry-adapted Hamiltonians, we are able to consider larger $r$ than for other, more complex molecules. To do this, we calculate the potential energy surface of molecular hydrogen in the STO-3G and STO-6G basis sets\cite{hehreSelfConsistentMolecularOrbitalMethods1969} for $1\leq r \leq 16$ for the SAE case. In Fig. \ref{fig_h2}, we present the results we obtained by diagonalizing the $r=16$ symmetry-adapted Ising Hamiltonian and compare them to the $r=4$ non-symmetry-adapted Ising Hamiltonian. We choose these values of $r$, as both cases require the same number of qubits (16) to simulate. We present the results calculated with the STO-3G basis set in Fig. \ref{fig_h2}, and present the results for the STO-6G, 3-21G\cite{binkleySelfconsistentMolecularOrbital1980}, 6-31G\cite{ditchfieldSelfConsistentMolecularOrbitalMethods1971}, and cc-pVDZ\cite{dunningGaussianBasisSets1989} basis sets in the supplementary material. Only vanishing differences exist for molecular hydrogen in the STO-3G and STO-6G basis sets, and that increasing $r$ is much more effective at reducing errors in the potential energy surface than increasing the size of the basis set.

At $r=16$, the symmetry-adapted ground-state energy curve matches the FCI result extremely well. Specifically in the regions $0.4 \textup{\AA} \leq R \leq 1.4$ $\textup{\AA}$ and $1.6 \textup{\AA} \leq R \leq 2.6$ $\textup{\AA}$, where $R$ is the interatomic distance of H$_2$, our symmetry-adapted method performs better than the original, non-symmetry-adapted method despite utilizing the same number of qubits to simulate the Hamiltonian. Because symmetry-adapted Ising Hamiltonian only contains $r$ qubits, while the non-symmetry-adapted Ising Hamiltonian contains $4r$ qubits, we are able to access a much larger value of $r$ for the SAE case (Table S1 of the supplementary material).

\begin{figure*}[t!]
    \centering
    \includegraphics[width=\textwidth]{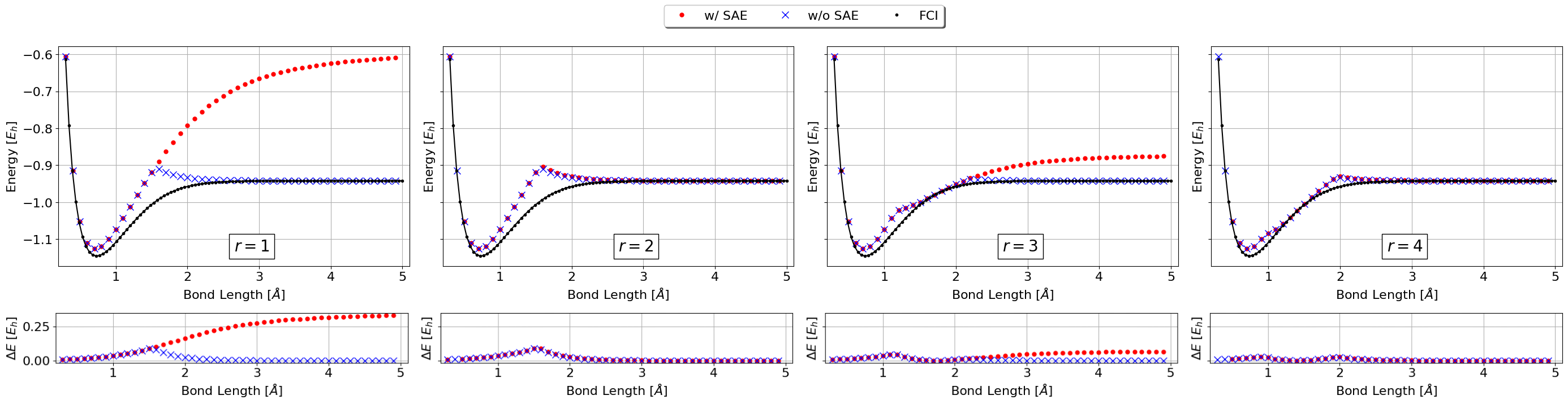}
    \caption{The potential energy surface of H$_2$ for $r\in\{1,2,3,4\}$ in STO-3G basis set. The red dots correspond to the symmetry-adapted Ising Hamiltonian ground-state energies, while the blue crosses correspond to the non-symmetry-adapted Ising Hamiltonian ground state eneriges. The black curve is the FCI ground state calculated in the STO-3G basis set, using PySCF. The bottom row shows the difference ($\Delta E$) between the symmetry-adapted or non-symmetry-adapted Ising Hamiltonian ground-state energy and the FCI ground-state energy.}
    \label{fig_h2_r}
\end{figure*}

We now dedicate the rest of the section to investigating the differences that emerge between the symmetry-adapted Ising Hamiltonian and the non-symmetry-adapted Ising Hamiltonian for molecular hydrogen. We find that these differences are most profound at low, odd values of the variational parameter $r$. To demonstrate this, we show results for the ground-state energy curves for $r\in\{1,2,3,4\}$ for both the symmetry-adapted and non-symmetry-adapted cases in Fig. \ref{fig_h2_r}. 

\begin{figure*}[t!]
    \centering
    \includegraphics[width=\textwidth]{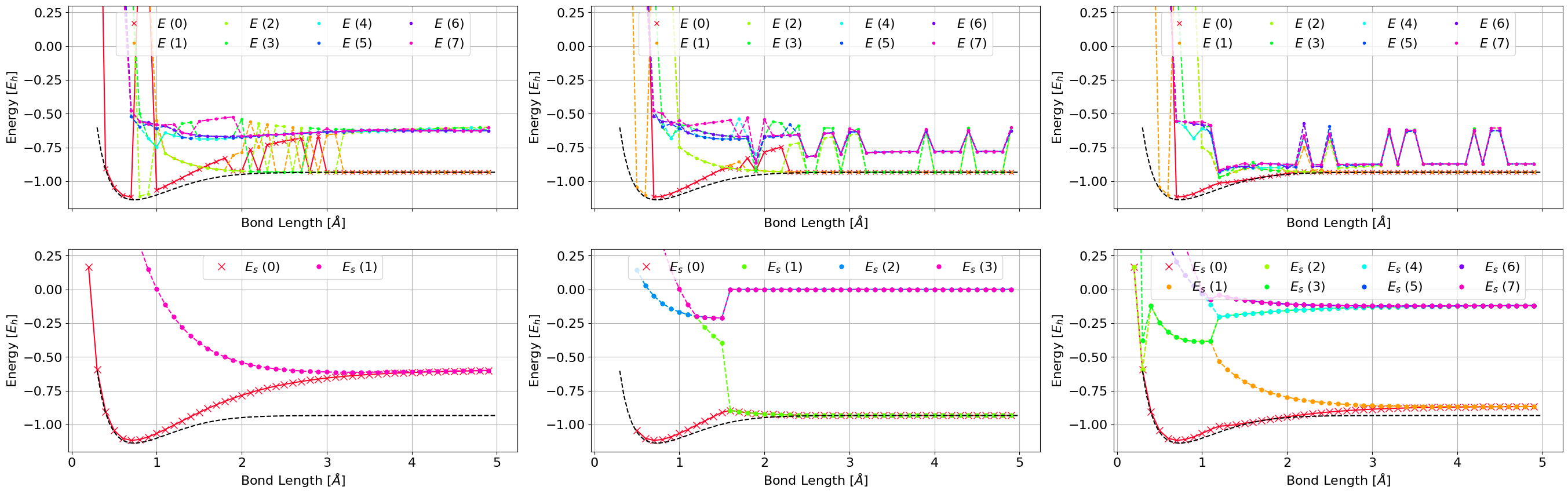}
    \caption{Top row: The first eight higher-order eigenvalues of the non-symmetry-adapted Ising Hamiltonian of H$_2$ for $r=1\text{ (leftmost)},\: 2 \text{ (center)}, \: 3 \text{ (rightmost)}$. Bottom row: the full eigenspectrum of the symmetry-adapted Ising Hamiltonian for H$_2$, matching the same $r$ value as the non-symmetry-adapted case. The black dashed line corresponds to the FCI ground-state energy.}
    \label{fig_h2_eig}
\end{figure*}

It can be seen from Fig. \ref{fig_h2_r} that there is a clear difference between the symmetry-adapted Ising Hamiltonian ground-state energy curve and the non-symmetry-adapted curve, especially profound for odd values of $r$. In particular for odd $r$, as we increase the bond length, we see that the symmetry-adapted curve converges to an incorrect energy value (approximately $-0.6$ Hartree for $r=1$ and approximately $-0.95$ Hartree for $r=3$). Furthermore, when comparing equal values of $r$, the non-symmetry-adapted Ising Hamiltonian performs consistently better than the symmetry-adapted Ising Hamiltonian. This behavior is unexpected, because as expected by the literature \cite{bravyiTaperingQubitsSimulate2017}$^,$\cite{setiaReducingQubitRequirements2020}$^,$\cite{picozziSymmetryadaptedEncodingsQubit2023}, the application of SAE should not affect the Hamiltonian ground-state energies. However, our results here show that the application of the XBK transformation to symmetry-adapted-encoded Hamiltonian necessarily degrades the ground-state energy curve for H$_2$. This explains the major differences seen for low, odd values of $r$ for H$_2$ in previous work that considered H$_2$ with a reduction of two qubits by symmetry in \cite{streifSolvingQuantumChemistry2019}.

We now quantify this degradation in terms of the full (or, extended) eigenspectrum of the XBK-transformed Hamiltonians. In general, an $m$ qubit Hamiltonian that is invariant under a $\mathbb{Z}_2^k$ symmetry will have $2^{rm}$ eigenvalues in its full Ising spectrum without SAE and will have $2^{r(m-k)}$. Hence, the application of SAE automatically removes $(1 - 2^{-kr}) 2^{mr}$ eigenvalues from consideration. We also point out that only the ground state of the XBK-transformed is a physically interpretable state \cite{xiaElectronicStructureCalculations2018} - the other remaining eigenvalues in the full Ising spectrum are internal, nonphysical "excited states," which we now refer to collectively as the higher-order eigenspectrum. In Fig. \ref{fig_h2_eig}, we present both the full higher-order eigenspectrum for the symmetry-adapted Ising Hamiltonian and the first eight higher-order eigenvalues for the non-symmetry-adapted Ising Hamiltonian for $r\in\{1,2,3\}$ for molecular hydrogen in STO-3G. The ground-state energies for each $r$ value correspond to the results presented in Fig. \ref{fig_h2_r}. 

At $r=1$, in the higher-order eigenspectrum for the non-symmetry-adapted Ising Hamiltonian, we observe a transition that occurs at $R=1.6 \textup{\AA}$ in where the first-excited eigenvalue $\lambda_1$ breaks its degeneracy with the second-excited eigenvalue $\lambda_2$ and becomes degenerate with the ground-state energy $E_0$ for the remainder of the bond lengths considered. Notably, such a feature is not present in the full higher-order eigenspectrum for the symmetry-adapted Ising Hamiltonian at $r=1$. This feature is again missed when considering $r=3$, as a gap persists for all values of the bond length and no such ground-state degeneracy occurs. This behavior is unlike, however, what occurs at $r=2$, in where the first higher-order eigenvalue of the symmetry-adapted Ising Hamiltonian $\lambda_1^{(s)}$ spontaneously drops to meet the ground-state energy curve $E_{0}^{(s)}$ and remains degenerate for all larger bond lengths considered. At $r=3$, we see a similar behavior as with $r=1$, except that the degeneracy of the ground state does not occur for the non-symmetry-adapted case until $R=2.2$ $\textup{\AA}$, yet does not occur for the symmetry-adapted case for all bond lengths considered.  

Hence, we claim that the application of the XBK transformation in coordination with a symmetry-adapted JW encoding removes the required higher-order eigenstates required to accurately model the ground state for odd $r$. Therefore, because the necessary states are missing, the lowest eigenvalue of the symmetry-adapted Ising Hamiltonian is confined to converge to an incorrect eigenvalue which corresponds to a higher-order eigenvalue of the non-symmetry-adapted Ising Hamiltonian.

\begin{figure*}[t!]
    \centering
    \includegraphics[width=0.75\textwidth]{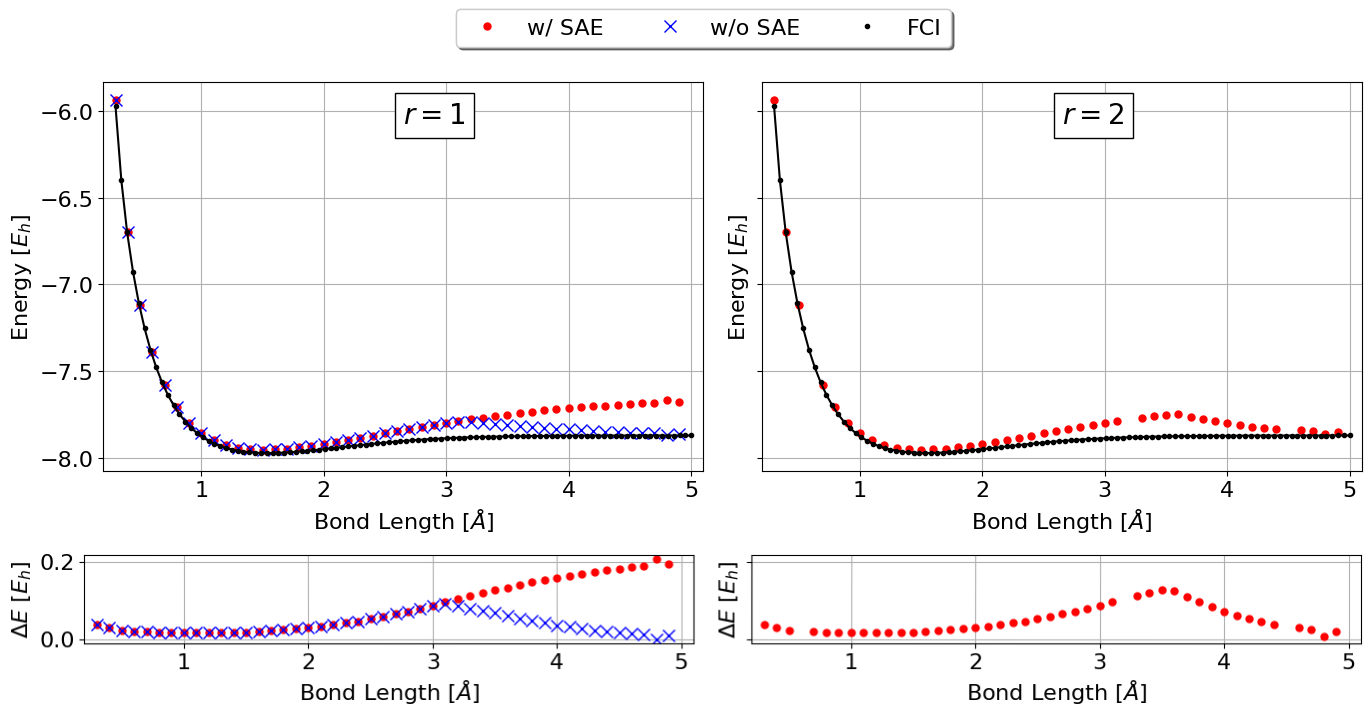}
    \caption{The potential energy surface of LiH for $r\in\{1,2\}$ in the STO-6G basis set.  The black curve is the FCI ground-state energy calculated in the STO-3G basis set, using PySCF. The bottom row shows the difference ($\Delta E$) between the symmetry-adapted or non-symmetry-adapted Ising Hamiltonian ground-state energy and the FCI ground-state energy. At $r=2$, we were only able to obtain the potential energy surface of the symmetry-adapted Ising Hamiltonian, as the non-symmetry-adapted case contained too many qubits to be diagonalized.}
    \label{fig_lih}
\end{figure*}

\begin{figure}[t!]
    \includegraphics[width=\columnwidth] {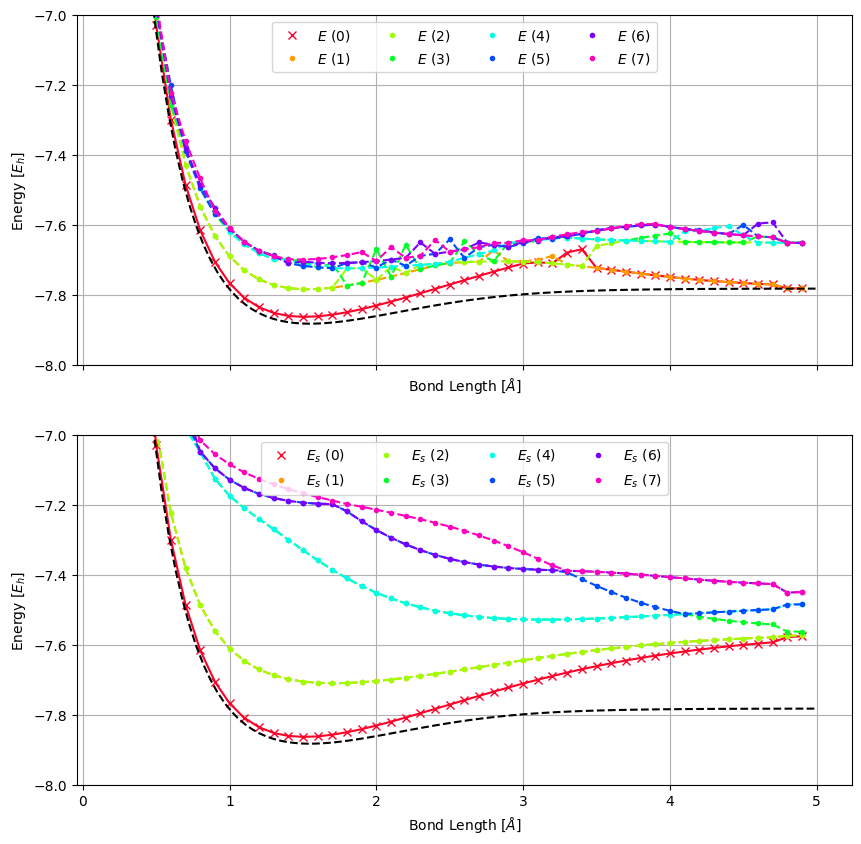}
    \caption{The first eight higher-order eigenvalues for both the non-symmetry-adapted (top) and symmetry-adapted (bottom) Ising Hamiltonians for LiH. We only present the results for $r=1$, as the LAPACK routines failed to converge for the 16-qubit LiH $r=2$ symmetry-adapted Ising Hamiltonian. The black dashed line is the FCI ground-state energy.}
    \label{fig_lih_eig}
\end{figure}

\subsubsection{Lithium Hydride - LiH}

As one of the simplest heteronuclear diatomic molecules, lithium hydride (LiH) is the second molecule that we consider. LiH requires twelve qubits without SAE and eight qubits with a symmetry-adapted JW encoding in a minimal basis set like STO-3G or STO-6G (see Table S1 of the supplementary material). The difference of four qubits arises because the largest Boolean symmetry group of LiH is $P^{\uparrow} \times P^{\downarrow} \times \sigma_{v}(xy) \times \sigma_{v}(yz) \cong \mathbb{Z}_2^4$, where $\sigma_{v}(\alpha \beta)$ corresponds to the unique reflection symmetries about the $\alpha \beta$ plane for $\alpha,\beta \in \{x,y,z\}$. The number of qubits for LiH is larger than for H$_2$, but not too large to be intractable for modern methods, including exact diagonalization \cite{xiaElectronicStructureCalculations2018} and implementation on a quantum annealing device \cite{streifSolvingQuantumChemistry2019}. 

In Fig. \ref{fig_lih}, we present our results for LiH in the STO-6G basis set. We only consider $r = 1$ for the non-symmetry-adapted case, as it is currently intractable to diagonalize a $12 \times 2 = 24$-qubit problem on classical hardware due to large memory requirements. We do, however, consider $r=2$ for the symmetry-adapted case as it is possible to diagonalize a $8 \times 2 = 16$-qubit problem. We note that there are some missing values on the potential energy curve at $r=2$ - these arise due to random issues with memory usage and convergence of the IRLM.

For LiH, we see a similar behavior to H$_2$ for intermediate and large bond lengths - specifically, we see that the symmetry-adapted potential energy curve diverges to an incorrect energy value for bond lengths greater than $R=3 \textup{\AA}$ for $r=1$ and that the potential energy curve converges to the correct energy value for $r=2$. We expect a similar behavior to extend to higher odd/even values of $r$, where for odd $r$, the symmetry-adapted Ising energy curve converges to an incorrect energy, while for even $r$ the symmetry-adapted Ising energy curve will converge to the FCI value.

The similarities of LiH to H$_2$ are also apparent in Fig. \ref{fig_lih_eig}, where we present the eight smallest higher-order eigenvalues for $r=1$ for both the symmetry-adapted and non-symmetry-adapted Ising Hamiltonians. We see that in the non-symmetry-adapted case at approximately $R=3 \textup{\AA}$, the first higher-order eigenvalue $\lambda_1$ breaks its degeneracy with $\lambda_2$ and becomes degenerate with the ground-state energy $E_0$. This effect is not present in the symmetry-adapted case. Instead, for the symmetry-adapted case, as seen with H$_2$, our ground-state energy $E_{0}^{(s)}$ converges to the the same value as a higher-order eigenvalue in the non-symmetry-adapted case for large bond lengths. This clearly signifies that the necessary higher-order state is again not present in the $r=1$ symmetry-adapted Ising eigenspectrum. However, for $r=2$ in Fig. \ref{fig_lih}, we do see that the ground-state energy $E_{0}^{(s)}$ converges to the FCI ground-state energy for larger bond lengths, signifying that the necessary higher-order eigenvalue is present. 

\subsubsection{The Helium Dimer - He$_2$}

The helium dimer, He$_2$, is another small homonuclear molecule that maintains the same symmetries as H$_2$. Due to this, in a minimal basis like STO-3G and STO-6G, we only require one qubit to represent the symmetry-adapted JW Hamiltonian. Therefore, we are able to extensively test our method as it scales with $r$, providing a comparison of the Ising Hamiltonian with and witho1ut SAE up to $r=4$.

\begin{figure*}[t!]
    \centering
    \includegraphics[width=\textwidth]{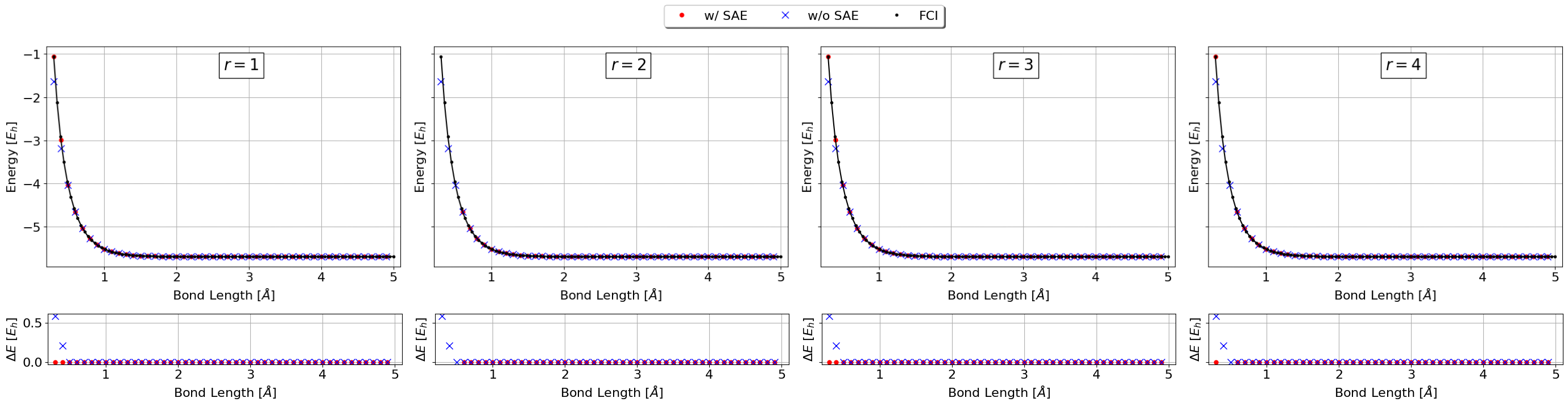}
    \caption{The potential energy surface of He$_2$ for $r\in\{1,2,3,4\}$ in the STO-6G basis set. }
    \label{fig_he2}
\end{figure*}

We present the results of our algorithm applied to He$_2$ in Fig. \ref{fig_he2} in the STO-6G basis set. Unlike H$_2$ and LiH, we see that no difference appears between the symmetry-adapted and non-symmery-adapted cases for large bond lengths. In fact, the only minor difference we observe is that the symmetry-adapted case is able to more accurately model the small bond length (<$0.5\textup{\AA}$) limit. Furthermore, we see that increasing $r$ leads to exponentially small reductions in the error $\Delta E$ and therefore only $r=1$ is required for highly accurate simulation of the potential energy curve of He$_2$. This implies that only a single-qubit Hamiltonian, the $r=1$ symmetry-adapted Ising Hamiltonian, is sufficient to model the potential energy curve of He$_2$ to within chemical accuracy. This drastically differs from the $r=1$ symmetry-adapted case for H$_2$, where upward of 16 qubits ($r=16$) are required to model the potential energy curve of H$_2$ to within chemical accuracy.

\begin{figure*}[t!]
    \centering
    \includegraphics[width=\textwidth]{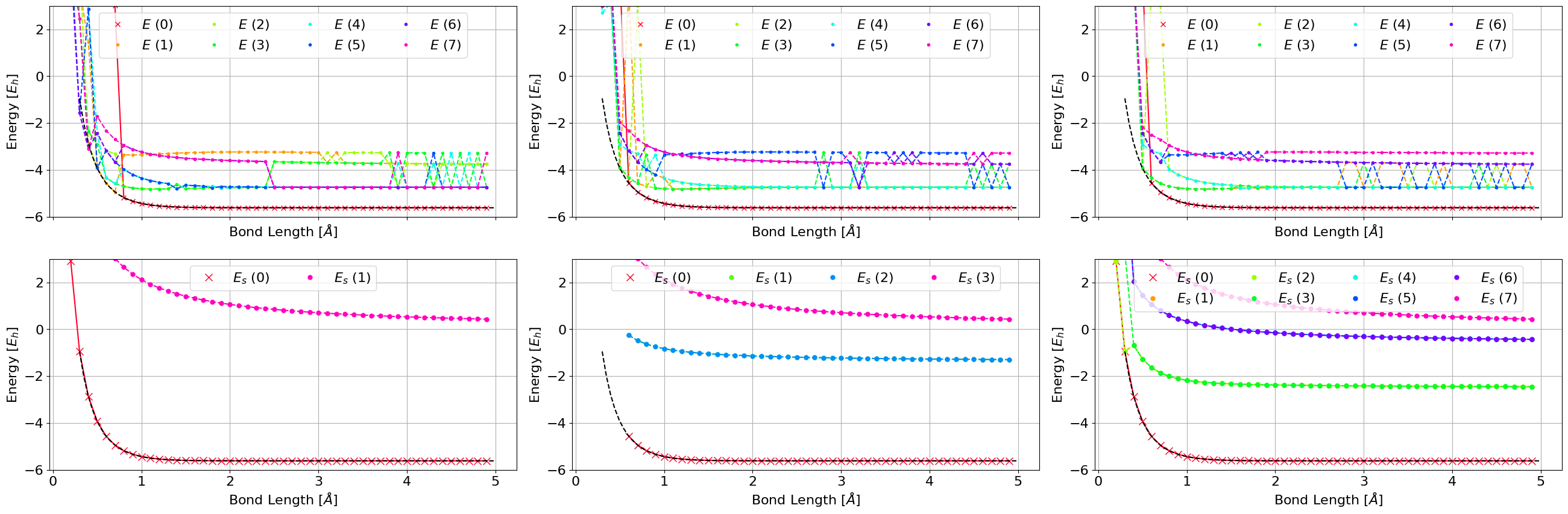}
    \caption{The higher-order eigenspectrum for He$_2$ for $r=1\text{ (leftmost)},\: 2 \text{ (center)}, \: 3 \text{ (rightmost)}$. Top row: The first eight non-symmetry-adapted higher-order eigenvalues. Bottom row: The full higher-order eigenspectrum for the symmetry-adapted Ising Hamiltonian. The black dashed line is the FCI ground-state energy.}
    \label{fig_he2_eig}
\end{figure*}

The reason why $r=1$ suffices for He$_2$ can be seen from Fig. \ref{fig_he2_eig}, as the full eigenspectrum for both the symmetry-adapted and non-symmetry-adapted Ising Hamiltonians are qualitatively very different from those of both H$_2$ and LiH. In fact, for the symmetry-adapted Ising Hamiltonian and at all $r\in\{1,2,3\}$, $E_{0}^{(s)}$ is gapped from $\lambda_1$ for all bond lengths considered. In fact, unlike H$_2$ and LiH, it is the first higher-order eigenvalue $\lambda_1$ that experiences a break and a reordering of its degeneracies. Because this level is consistently gapped from the ground state for all intermediate to large values of $R$, the ground state does not cross with any higher-order eigenvalues and thus cannot converge to an incorrect higher-order eigenvalue of the symmetry-adapted case.

\subsubsection{The Water Molecule - H$_2$O}

The water molecule with the XBK method has also been simulated, however, this time only using simulated and quantum annealing as presented in Copenhaver et al.\cite{copenhaverUsingQuantumAnnealers2021} As the water molecule is larger than molecules previously considered, requiring 14 qubits to represent the full qubit Hamiltonian in a minimal basis (see Table S1 of the supplementary material), simulation of the potential energy surface without active space restrictions has not yet been performed on a quantum annealing device or with the XBK method \cite{copenhaverUsingQuantumAnnealers2021}. Therefore, since we do not perform any active space restrictions, as far as we know this is the first time the full ten-electron potential energy surface of H$_2$O has been reported to be simulated with a method designed for quantum annealers that approaches the FCI accuracy. To create the symmetry-adapted JW Hamiltonian for H$_2$O, we note that the largest Boolean symmetry group of the molecule is $P^{\uparrow} \times P^{\downarrow} \times \sigma_{v}(yz) \times \sigma_{v}(xz) \cong \mathbb{Z}_2^4$. This then enables us to rewrite the electronic Hamiltonian with only ten qubits.

\begin{figure}[hbt]
    \includegraphics[width=\columnwidth] {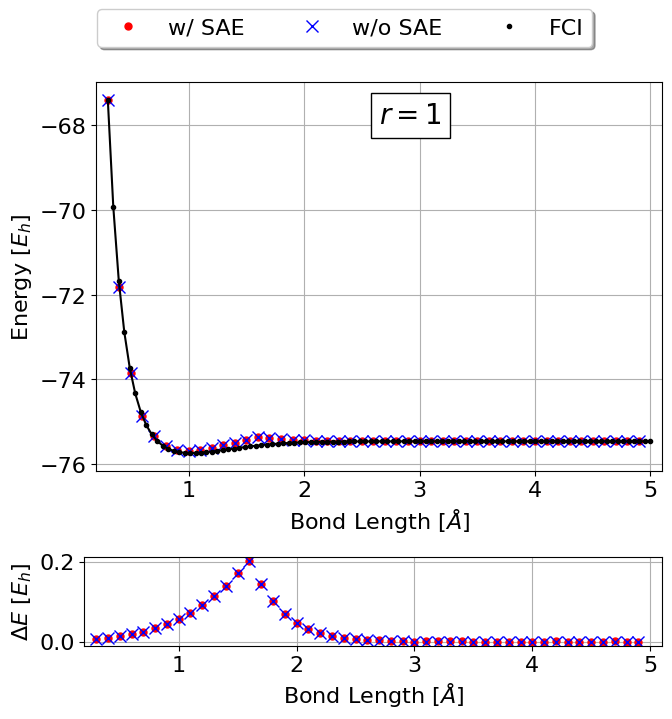}
    \caption{The potential energy surface of H$_2$O at $r=1$ in the STO-6G basis set. $\Delta E$ is the difference between the symmetry-adapted or non-symmetry-adapted Ising Hamiltonian ground-state energy and the FCI ground-state energy.}
    \label{fig_h2o}
\end{figure}

We present the potential energy surface for H$_2$O at $r=1$ in the STO-6G basis set in Fig. \ref{fig_h2o}, where we symmetrically vary the O-H bond length. The H-O-H bond angle is kept at 104.5\degree. Similar to He$_2$, at $r=1$ there is no significant difference between the non-symmetry-adapted and symmetry-adapted Ising ground-state curves. Unlike He$_2$, however, we see that $r=1$ is insufficient to model the potential energy curve to within chemical accuracy for either case. In order to increase the accuracy of our method, it is straightforward to increase $r$. However, with current computational constraints, we are currently unable to diagonalize a $12\times2=24$- or $10\times2=20$-qubit problem, and thus the use of higher $r$ for simulation is beyond reach due to contemporary memory and time constraints of classical hardware. 

We note that because the symmetry-adapted Ising ground state does not differ from the non-symmetry-adapted case, H$_2$O behaves much more similarly to He$_2$ than to H$_2$ and LiH. 

\begin{figure}[hbt]
    \includegraphics[width=\columnwidth] {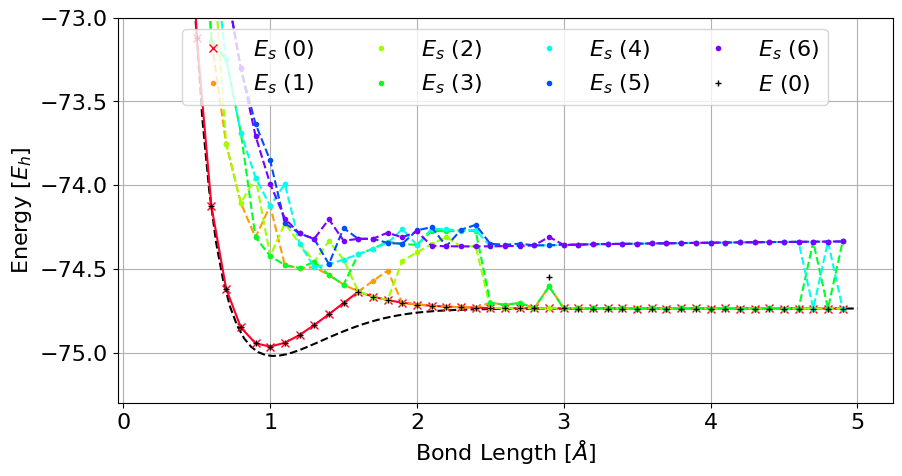}
    \caption{The first seven higher-order eigenvalues for the symmetry-adapted $r=1$ Ising Hamiltonian for H$_2$O, overlaid with the FCI ground-state energy (black dashed line) and the ground state of the non-symmetry-adapted $E(0)$.}
    \label{fig_h2o_eig}
\end{figure}

In Fig. \ref{fig_h2o_eig}, we present the first eight eigenvalues of the higher-order eigenspectrum for the symmetry-adapted Ising Hamiltonian of H$_2$O together with the non-symmetry-adapted ground-state energy curve. We do not present the first eight eigenvalues of the higher-order eigenspectrum for the non-symmetry-adapted Ising Hamiltonian as the classical calculations failed to converge, and instead only present the ground state. We see that, similar to He$_2$, the ground-state energy of the symmetry-adapted Ising Hamiltonian becomes degenerate with the first higher-order eigenvalue from the FCI curve at large bond lengths. Furthermore, we see that the doubly-degenerate ground state of the symmetry-adapted Ising Hamiltonian lies along the same curve as the non-symmetry-adapted ground state. Therefore, similar to He$_2$ and unlike LiH and H$_2$, we see that the required states of the symmetry-adapted Ising Hamiltonian are present and match the non-symmetry-adapted ground state. Hence, symmetry-adapted encodings do not remove the critical higher-order states to model the ground state of H$_2$O.

\subsubsection{Borane - BH$_3$}

To the best of our knowledge, this is the first time that an energy surface of borane (BH$_3$) is obtained using the XBK method. We simulate the potential energy surface by symmetrically varying the three B-H bonds, as we did with H$_2$O, with a H-B-H bond angle of 120\degree. Borane requires 16 qubits to be simulated without a SAE in a minimal basis set, and can be simplified to 12 qubits owning to its $P^{\uparrow} \times P^{\downarrow} \times \sigma_{v}(yz) \times \sigma_{v}(xz) \cong \mathbb{Z}_2^4$ symmetry. Therefore, like H$_2$O, our analysis is limited to only including $r=1$.

\begin{figure}[hbt]
    \includegraphics[width=\columnwidth] {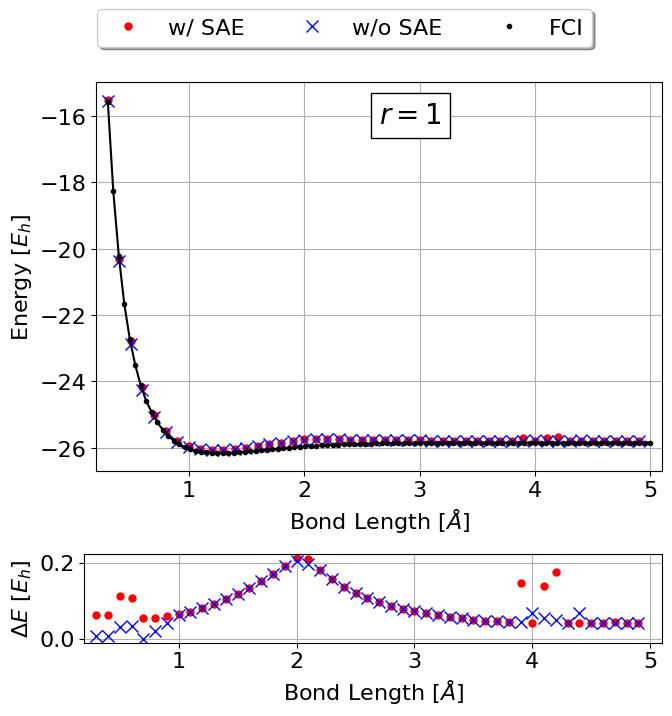}
    \caption{The potential energy surface of borane (BH$_3$) at $r=1$ in the STO-6G basis set. $\Delta E$ is the difference between the symmetry-adapted or non-symmetry-adapted Ising Hamiltonian ground-state energy and the FCI ground-state energy.}
    \label{fig_bh3}
\end{figure}

We present the potential energy surface for borane in Fig. \ref{fig_bh3} using the STO-6G basis set. We note that the classical FCI method we were using to construct the baseline curves struggles to converge at all bond lengths considered, even in a minimal basis set. Therefore, for some bond lengths considered in $R \in [0.1,5.0] \textup{\AA}$, we are missing the FCI value. Because only a very small number of points are missing, we use linear interpolation to estimate those values.

\begin{figure}[hbt]
    \includegraphics[width=\columnwidth] {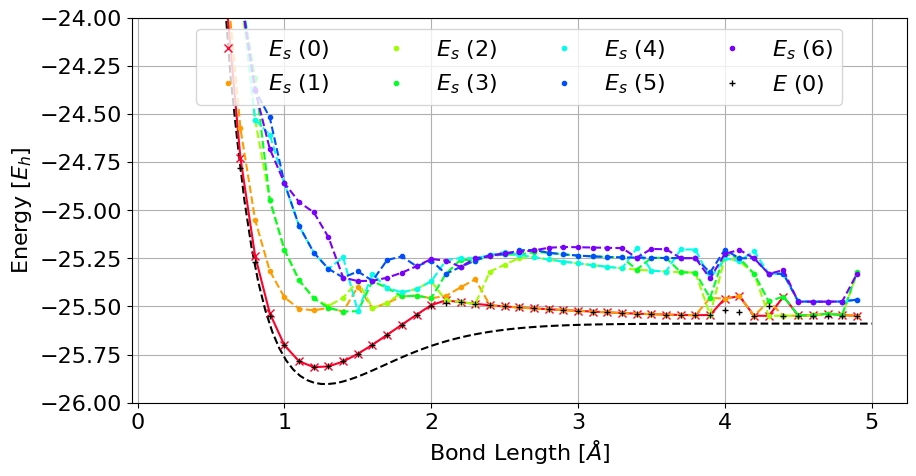}
    \caption{The first seven higher-order eigenvalues for the symmetry-adapted $r=1$ Ising Hamiltonian for BH$_3$, overlaid with the FCI ground-state energy (black dashed line) and the ground state of the non-symmetry-adapted $E(0)$.}
    \label{fig_bh3_eig}
\end{figure}

We note that the symmetry-adapted curve is meaningfully different (difference greater than 0.1 Hartree) from the non-symmetry-adapted curve in two regions: first, in the low bond-length limit $R<1.0 \textup{\AA}$, and second in the range $4.0 \textup{\AA}\leq R \leq4.4 \textup{\AA}$. 
We can see from Fig. \ref{fig_bh3_eig} that there are only spontaneous spots in which the symmetry-adapted ground-state energy "jumps" from the non-symmetry-adapted ground-state energy, which corresponds exactly to the errors we see in Fig. \ref{fig_bh3}. Similar to H$_2$ and LiH, we see that the ground-state energy of the symmetry-adapted Hamiltonian jumps to meet an incorrect state for these regions. However, unlike H$_2$ and LiH, we see that the symmetry-adapted ground state correctly converges to the FCI ground state in the large bond-length limit, as the large ground-state degeneracy occurs near 5 $\textup{\AA}$, signifying that a large number of necessary states persist in this limit. 

\subsubsection{Ammonia - NH$_3$}

The final molecule that we consider here in which both symmetry-adapted and non-symmetry-adapted Ising Hamiltonian ground-state energy curves were obtained is ammonia (NH$_3$). We calculate the energy surface similarly to those of H$_2$O and BH$_3$, varying each N-H bond symmetrically with a bond angle of 107.8\degree. To the best of our knowledge, ammonia, which requires 16 qubits to simulate without SAE and 13 qubits with (due to ammonia's $P^\uparrow \times P^\downarrow \times \sigma_h \cong \mathbb{Z}_2^3$ symmetry, where $\sigma_h$ corresponds to a reflection across the horizontal plane) in a minimal basis set, is the largest molecule simulated with the XBK method. Therefore, this molecule is the best test of the scaling capability of both the symmetry-adapted and non-symmetry-adpated XBK methods. In the supplementary material, we do however present the results of molecules that require more qubits to simulate than ammonia, including N$_2$, O$_2$, CO, F$_2$, Li$_2$, and CH$_4$, up to 16 qubits with SAE. We do not present these results here, though, as the non-symmetry-adapted cases require too many qubits to be reasonably simulated on modern classical or quantum hardware and their comparison with the SAE results cannot be done. 

\begin{figure}[hbt]
    \includegraphics[width=\columnwidth] {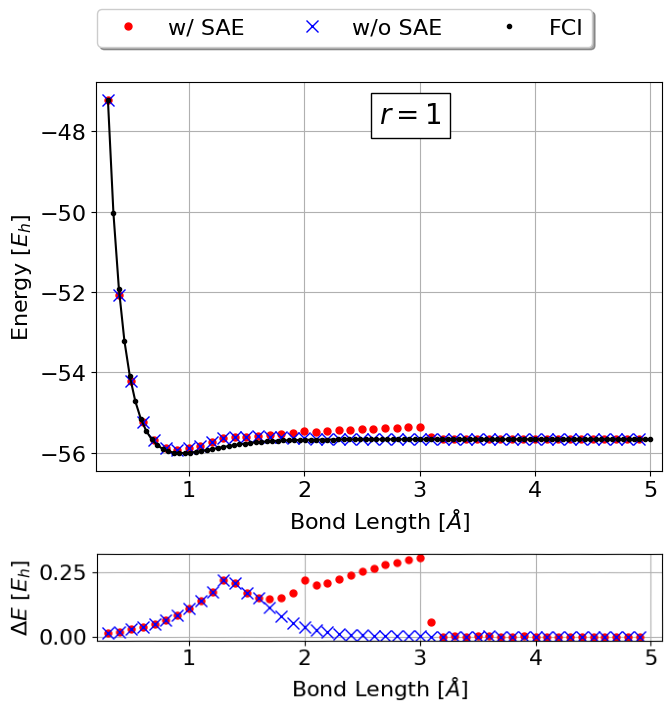}
    \caption{The potential energy surface of NH$_3$ at $r=1$ in the STO-6G basis set. $\Delta E$ is the difference between the symmetry-adapted or non-symmetry-adapted Ising Hamiltonian ground-state energy and the FCI ground-state energy.}
    \label{fig_nh3}
\end{figure}

We present our results for the potential energy surface of ammonia in Fig. \ref{fig_nh3}.  Like BH$_3$, due to the size of NH$_3$, not all of our FCI calculations converge. The results for NH$_3$ are qualitatively similar to those of H$_2$, LiH, and BH$_3$, as the symmetry-adapted case differs from the non-symmetry-adapted case. However, the behavior of the symmetry-adapted Ising Hamiltonian potential energy surface of NH$_3$ differs from these examples specifically because after approximately 3.0 $\textup{\AA}$, we see that the symmetry-adapted ground state spontaneously drops to meet the non-symmetry-adapted and FCI ground states. 

\begin{figure}[hbt]
    \includegraphics[width=\columnwidth] {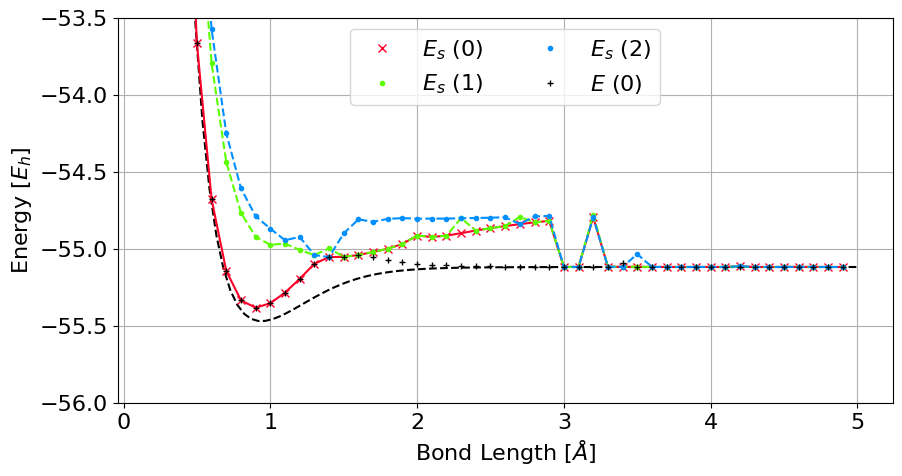}
    \caption{The first three higher-order eigenvalues for the symmetry-adapted $r=1$ Ising Hamiltonian for NH$_3$, overlaid with the FCI ground-state energy (black dashed line) and the ground state of the non-symmetry-adapted $E(0)$.}
    \label{fig_nh3_eig}
\end{figure}

From Fig. \ref{fig_nh3_eig}, we can see that in the main region, the symmetry-adapted curve differs from the non-symmetry-adapted curve because the symmetry-adapted Ising Hamiltonian does not possess the relevant converging state between approximately 1.6 $\textup{\AA}$ and 3.0 $\textup{\AA}$. Therefore, the ground-state energy of the symmetry-adapted Ising Hamiltonian is not able to relax into the proper ground state in this region, and is thus confined to relax into a higher-order eigenvalue. We see this explicitly as the first three higher-order eigenvalues of the symmetry-adapted case converge to the same incorrect value at 3.0 $\textup{\AA}$. Therefore, we see again the behavior in which the symmetry-adapted encodings remove the relevant states to properly model the ground state, beyond the errors collected by the non-symmetry-adapted case.

\section{Discussion}

We notice that for molecules where the ground state of the symmetry-adapted Ising Hamiltonian significantly differs from the ground state of the non-symmetry-adapted Ising Hamiltonian, including H$_2$, LiH, BH$_3$, and NH$_3$, at the dissociation limit they consist of only radicals. As the bonds stretch symmetrically, the electronic configuration of the system becomes frustrated. No single electronic configuration can fully describe the system in this intermediate interatomic distance region, necessitating a multiconfigurational approach to accurately represent the electronic structure. Since the symmetry-adapted JW encoding necessarily eliminates $(1 - 2^{-kr}) 2^{mr}$ internally excited states from consideration of the XBK-transformed Ising Hamiltonian, we conclude that specifically for odd values of $r$ the relevant states required to properly model the electronic structure lie within the $(1 - 2^{-kr}) 2^{mr}$ removed states. We can explicitly see this by considering the higher-order eigenspectrum of the symmetry-adapted Ising Hamiltonian, e.g., Fig. \ref{fig_h2_eig}, as there does not exist the relevant state(s) present in the non-symmetry-adapted Ising Hamiltonian higher-order spectrum needed to adiabatically relax into the true ground state.

For H$_2$O and He$_2$, the difference between the symmetry-adapted and non-symmetry-adapted Ising Hamiltonian ground-state energies is non-significant. This is because, in these systems, no necessary eigenstates in the extended eigenspectrum to model the ground state are present in the $(1 - 2^{-kr}) 2^{mr}$ states that are removed when applying symmetry-adapted encodings with the XBK method. 

For bond dissociation that does not have a high multireference character, such as He$_2$, symmetry-adapted encodings can model the potential energy surface with the XBK method to FCI accuracy even for small $r$ values. The symmetric dissociation of H$_2$O is considered multiconfigurational, similar to H$_2$, LiH, BH$_3$, and NH$_3$, resulting in the error at the intermediate bond length compared to FCI for both the symmetry-adapted and the non-symmetry adapted Ising Hamiltonian ground-state energies. In this case, a large $r$ is necessary to model the bond dissociation curves at FCI accuracy for both the symmetry-adapted and non-symmetry-adapted XBK method. H$_2$O is a peculiar case as, despite of its multireference character along the bond dissociation curve, its results from SAE are of the same quality as the non-symmetry-adapted case, similar to He$_2$. 

For practical purposes, we note that for systems that do not have a high multireference character, one is able to apply symmetry-adapted encodings with small $r$ for the XBK transformation and suffer no additional errors. Therefore, when calculating potential energy surfaces for this class of molecules on quantum annealers, one should always apply symmetry-adapted encodings. For systems that do have a high multireference character, an even $r$ should be used to not suffer additional errors from the symmetry adaptation. When a higher accuracy is needed, one must balance the number of qubits saved from symmetry-adapted encodings with the additional errors attained by their application. For large $r$, these errors are vanishing, so if one is able to allocate a large number of qubits for their calculation and hence utilize a higher $r$ value, symmetry-adapted encodings should also always be used.  

\section{Conclusion} 

In this work, we have shown that our algorithm that combines symmetry-adapted encodings based on the full Boolean symmetry group with our implementation of the XBK algorithm allows the simulation of larger molecules than previously considered and without the need for active space restrictions (including specifically BH$_3$ and NH$_3$, as well as N$_2$, F$_2$, O$_2$, CO, Li$_2$, and CH$_4$ in the supplementary material). Furthermore, we provide an explanation for previously unexplained behavior\cite{streifSolvingQuantumChemistry2019} of the potential energy surfaces for low, odd values of $r$.

Our method scales particularly well for systems that require a large $r$ value to be properly simulated. This is because, for a molecular system with a $\mathbb{Z}_2^k$ symmetry, our method saves $rk$ qubits from consideration. Therefore, our method produces an exponential reduction in the size of the Hilbert space for the system, as the Hilbert space scales with $2^m$ for $m$ qubits. In the future, when larger-scale quantum annealing hardware becomes available, our method can be applied to make the simulation of large molecular systems currently unaffordable at FCI accuracy more realistic.

\section*{Supplementary Material}
Supplementary material containing the following has been provided: Equivalence of our implementation of the XBK method, effect of basis sets on the potential energy surface of H$_2$, potential energy surfaces for N$_2$, O$_2$, CO, F$_2$, Li$_2$, and CH$_4$, and qubit counts of the molecules discussed in this work. 

\begin{acknowledgments}
This work is supported by the U.S. Department of Energy, Office of Science, Office of Advanced Scientific Computing Research, under Award Number DE-SC0024216. J.D. acknowledges the Northeastern University Undergraduate Research and Fellowships PEAK Experiences Award. We thank Northeastern University Research Computing for providing computing resources. We also thank Dr. Kübra Yeter-Aydeniz for valuable discussions and insight related to this work. 
\end{acknowledgments}

\bibliography{references}

\begin{thebibliography}{28}%
\makeatletter
\providecommand \@ifxundefined [1]{%
 \@ifx{#1\undefined}
}%
\providecommand \@ifnum [1]{%
 \ifnum #1\expandafter \@firstoftwo
 \else \expandafter \@secondoftwo
 \fi
}%
\providecommand \@ifx [1]{%
 \ifx #1\expandafter \@firstoftwo
 \else \expandafter \@secondoftwo
 \fi
}%
\providecommand \natexlab [1]{#1}%
\providecommand \enquote  [1]{``#1''}%
\providecommand \bibnamefont  [1]{#1}%
\providecommand \bibfnamefont [1]{#1}%
\providecommand \citenamefont [1]{#1}%
\providecommand \href@noop [0]{\@secondoftwo}%
\providecommand \href [0]{\begingroup \@sanitize@url \@href}%
\providecommand \@href[1]{\@@startlink{#1}\@@href}%
\providecommand \@@href[1]{\endgroup#1\@@endlink}%
\providecommand \@sanitize@url [0]{\catcode `\\12\catcode `\$12\catcode `\&12\catcode `\#12\catcode `\^12\catcode `\_12\catcode `\%12\relax}%
\providecommand \@@startlink[1]{}%
\providecommand \@@endlink[0]{}%
\providecommand \url  [0]{\begingroup\@sanitize@url \@url }%
\providecommand \@url [1]{\endgroup\@href {#1}{\urlprefix }}%
\providecommand \urlprefix  [0]{URL }%
\providecommand \Eprint [0]{\href }%
\providecommand \doibase [0]{http://dx.doi.org/}%
\providecommand \selectlanguage [0]{\@gobble}%
\providecommand \bibinfo  [0]{\@secondoftwo}%
\providecommand \bibfield  [0]{\@secondoftwo}%
\providecommand \translation [1]{[#1]}%
\providecommand \BibitemOpen [0]{}%
\providecommand \bibitemStop [0]{}%
\providecommand \bibitemNoStop [0]{.\EOS\space}%
\providecommand \EOS [0]{\spacefactor3000\relax}%
\providecommand \BibitemShut  [1]{\csname bibitem#1\endcsname}%
\let\auto@bib@innerbib\@empty
\bibitem [{\citenamefont {Feynman}(1982)}]{feynmanSimulatingPhysicsComputers1982}%
  \BibitemOpen
  \bibfield  {author} {\bibinfo {author} {\bibfnamefont {R.~P.}\ \bibnamefont {Feynman}},\ }\href {\doibase 10.1007/BF02650179} {\bibfield  {journal} {\bibinfo  {journal} {International Journal of Theoretical Physics}\ }\textbf {\bibinfo {volume} {21}},\ \bibinfo {pages} {467} (\bibinfo {year} {1982})}\BibitemShut {NoStop}%
\bibitem [{\citenamefont {McArdle}\ \emph {et~al.}(2020)\citenamefont {McArdle}, \citenamefont {Endo}, \citenamefont {{Aspuru-Guzik}}, \citenamefont {Benjamin},\ and\ \citenamefont {Yuan}}]{mcardleQuantumComputationalChemistry2020}%
  \BibitemOpen
  \bibfield  {author} {\bibinfo {author} {\bibfnamefont {S.}~\bibnamefont {McArdle}}, \bibinfo {author} {\bibfnamefont {S.}~\bibnamefont {Endo}}, \bibinfo {author} {\bibfnamefont {A.}~\bibnamefont {{Aspuru-Guzik}}}, \bibinfo {author} {\bibfnamefont {S.~C.}\ \bibnamefont {Benjamin}}, \ and\ \bibinfo {author} {\bibfnamefont {X.}~\bibnamefont {Yuan}},\ }\href {\doibase 10.1103/RevModPhys.92.015003} {\bibfield  {journal} {\bibinfo  {journal} {Reviews of Modern Physics}\ }\textbf {\bibinfo {volume} {92}},\ \bibinfo {pages} {015003} (\bibinfo {year} {2020})}\BibitemShut {NoStop}%
\bibitem [{\citenamefont {Nielsen}\ and\ \citenamefont {Chuang}(2010)}]{nielsenQuantumComputationQuantum2010}%
  \BibitemOpen
  \bibfield  {author} {\bibinfo {author} {\bibfnamefont {M.~A.}\ \bibnamefont {Nielsen}}\ and\ \bibinfo {author} {\bibfnamefont {I.~L.}\ \bibnamefont {Chuang}},\ }\href {\doibase 10.1017/CBO9780511976667} {\enquote {\bibinfo {title} {Quantum {{Computation}} and {{Quantum Information}}: 10th {{Anniversary Edition}}},}\ }\bibinfo {howpublished} {https://www.cambridge.org/highereducation/books/quantum-computation-and-quantum-information/01E10196D0A682A6AEFFEA52D53BE9AE} (\bibinfo {year} {2010})\BibitemShut {NoStop}%
\bibitem [{\citenamefont {Peruzzo}\ \emph {et~al.}(2014)\citenamefont {Peruzzo}, \citenamefont {McClean}, \citenamefont {Shadbolt}, \citenamefont {Yung}, \citenamefont {Zhou}, \citenamefont {Love}, \citenamefont {{Aspuru-Guzik}},\ and\ \citenamefont {O'Brien}}]{peruzzoVariationalEigenvalueSolver2014}%
  \BibitemOpen
  \bibfield  {author} {\bibinfo {author} {\bibfnamefont {A.}~\bibnamefont {Peruzzo}}, \bibinfo {author} {\bibfnamefont {J.}~\bibnamefont {McClean}}, \bibinfo {author} {\bibfnamefont {P.}~\bibnamefont {Shadbolt}}, \bibinfo {author} {\bibfnamefont {M.-H.}\ \bibnamefont {Yung}}, \bibinfo {author} {\bibfnamefont {X.-Q.}\ \bibnamefont {Zhou}}, \bibinfo {author} {\bibfnamefont {P.~J.}\ \bibnamefont {Love}}, \bibinfo {author} {\bibfnamefont {A.}~\bibnamefont {{Aspuru-Guzik}}}, \ and\ \bibinfo {author} {\bibfnamefont {J.~L.}\ \bibnamefont {O'Brien}},\ }\href {\doibase 10.1038/ncomms5213} {\bibfield  {journal} {\bibinfo  {journal} {Nature Communications}\ }\textbf {\bibinfo {volume} {5}},\ \bibinfo {pages} {4213} (\bibinfo {year} {2014})}\BibitemShut {NoStop}%
\bibitem [{\citenamefont {Nakanishi}\ \emph {et~al.}(2019)\citenamefont {Nakanishi}, \citenamefont {Mitarai},\ and\ \citenamefont {Fujii}}]{nakanishiSubspacesearchVariationalQuantum2019}%
  \BibitemOpen
  \bibfield  {author} {\bibinfo {author} {\bibfnamefont {K.~M.}\ \bibnamefont {Nakanishi}}, \bibinfo {author} {\bibfnamefont {K.}~\bibnamefont {Mitarai}}, \ and\ \bibinfo {author} {\bibfnamefont {K.}~\bibnamefont {Fujii}},\ }\href {\doibase 10.1103/PhysRevResearch.1.033062} {\bibfield  {journal} {\bibinfo  {journal} {Physical Review Research}\ }\textbf {\bibinfo {volume} {1}},\ \bibinfo {pages} {033062} (\bibinfo {year} {2019})}\BibitemShut {NoStop}%
\bibitem [{\citenamefont {Grimsley}\ \emph {et~al.}(2019)\citenamefont {Grimsley}, \citenamefont {Economou}, \citenamefont {Barnes},\ and\ \citenamefont {Mayhall}}]{grimsleyAdaptiveVariationalAlgorithm2019}%
  \BibitemOpen
  \bibfield  {author} {\bibinfo {author} {\bibfnamefont {H.~R.}\ \bibnamefont {Grimsley}}, \bibinfo {author} {\bibfnamefont {S.~E.}\ \bibnamefont {Economou}}, \bibinfo {author} {\bibfnamefont {E.}~\bibnamefont {Barnes}}, \ and\ \bibinfo {author} {\bibfnamefont {N.~J.}\ \bibnamefont {Mayhall}},\ }\href {\doibase 10.1038/s41467-019-10988-2} {\bibfield  {journal} {\bibinfo  {journal} {Nature Communications}\ }\textbf {\bibinfo {volume} {10}},\ \bibinfo {pages} {3007} (\bibinfo {year} {2019})}\BibitemShut {NoStop}%
\bibitem [{\citenamefont {Aharonov}\ \emph {et~al.}(2005)\citenamefont {Aharonov}, \citenamefont {{van Dam}}, \citenamefont {Kempe}, \citenamefont {Landau}, \citenamefont {Lloyd},\ and\ \citenamefont {Regev}}]{aharonovAdiabaticQuantumComputation2005}%
  \BibitemOpen
  \bibfield  {author} {\bibinfo {author} {\bibfnamefont {D.}~\bibnamefont {Aharonov}}, \bibinfo {author} {\bibfnamefont {W.}~\bibnamefont {{van Dam}}}, \bibinfo {author} {\bibfnamefont {J.}~\bibnamefont {Kempe}}, \bibinfo {author} {\bibfnamefont {Z.}~\bibnamefont {Landau}}, \bibinfo {author} {\bibfnamefont {S.}~\bibnamefont {Lloyd}}, \ and\ \bibinfo {author} {\bibfnamefont {O.}~\bibnamefont {Regev}},\ }\href {\doibase 10.48550/arXiv.quant-ph/0405098} {\enquote {\bibinfo {title} {Adiabatic {{Quantum Computation}} is {{Equivalent}} to {{Standard Quantum Computation}}},}\ } (\bibinfo {year} {2005}),\ \Eprint {http://arxiv.org/abs/quant-ph/0405098} {arXiv:quant-ph/0405098} \BibitemShut {NoStop}%
\bibitem [{\citenamefont {King}\ \emph {et~al.}(2023)\citenamefont {King}, \citenamefont {Raymond}, \citenamefont {Lanting}, \citenamefont {Harris}, \citenamefont {Zucca}, \citenamefont {Altomare}, \citenamefont {Berkley}, \citenamefont {Boothby}, \citenamefont {Ejtemaee}, \citenamefont {Enderud}, \citenamefont {Hoskinson}, \citenamefont {Huang}, \citenamefont {Ladizinsky}, \citenamefont {MacDonald}, \citenamefont {Marsden}, \citenamefont {Molavi}, \citenamefont {Oh}, \citenamefont {{Poulin-Lamarre}}, \citenamefont {Reis}, \citenamefont {Rich}, \citenamefont {Sato}, \citenamefont {Tsai}, \citenamefont {Volkmann}, \citenamefont {Whittaker}, \citenamefont {Yao}, \citenamefont {Sandvik},\ and\ \citenamefont {Amin}}]{kingQuantumCriticalDynamics2023}%
  \BibitemOpen
  \bibfield  {author} {\bibinfo {author} {\bibfnamefont {A.~D.}\ \bibnamefont {King}}, \bibinfo {author} {\bibfnamefont {J.}~\bibnamefont {Raymond}}, \bibinfo {author} {\bibfnamefont {T.}~\bibnamefont {Lanting}}, \bibinfo {author} {\bibfnamefont {R.}~\bibnamefont {Harris}}, \bibinfo {author} {\bibfnamefont {A.}~\bibnamefont {Zucca}}, \bibinfo {author} {\bibfnamefont {F.}~\bibnamefont {Altomare}}, \bibinfo {author} {\bibfnamefont {A.~J.}\ \bibnamefont {Berkley}}, \bibinfo {author} {\bibfnamefont {K.}~\bibnamefont {Boothby}}, \bibinfo {author} {\bibfnamefont {S.}~\bibnamefont {Ejtemaee}}, \bibinfo {author} {\bibfnamefont {C.}~\bibnamefont {Enderud}}, \bibinfo {author} {\bibfnamefont {E.}~\bibnamefont {Hoskinson}}, \bibinfo {author} {\bibfnamefont {S.}~\bibnamefont {Huang}}, \bibinfo {author} {\bibfnamefont {E.}~\bibnamefont {Ladizinsky}}, \bibinfo {author} {\bibfnamefont {A.~J.~R.}\ \bibnamefont {MacDonald}}, \bibinfo {author} {\bibfnamefont {G.}~\bibnamefont {Marsden}}, \bibinfo {author} {\bibfnamefont
  {R.}~\bibnamefont {Molavi}}, \bibinfo {author} {\bibfnamefont {T.}~\bibnamefont {Oh}}, \bibinfo {author} {\bibfnamefont {G.}~\bibnamefont {{Poulin-Lamarre}}}, \bibinfo {author} {\bibfnamefont {M.}~\bibnamefont {Reis}}, \bibinfo {author} {\bibfnamefont {C.}~\bibnamefont {Rich}}, \bibinfo {author} {\bibfnamefont {Y.}~\bibnamefont {Sato}}, \bibinfo {author} {\bibfnamefont {N.}~\bibnamefont {Tsai}}, \bibinfo {author} {\bibfnamefont {M.}~\bibnamefont {Volkmann}}, \bibinfo {author} {\bibfnamefont {J.~D.}\ \bibnamefont {Whittaker}}, \bibinfo {author} {\bibfnamefont {J.}~\bibnamefont {Yao}}, \bibinfo {author} {\bibfnamefont {A.~W.}\ \bibnamefont {Sandvik}}, \ and\ \bibinfo {author} {\bibfnamefont {M.~H.}\ \bibnamefont {Amin}},\ }\href {\doibase 10.1038/s41586-023-05867-2} {\bibfield  {journal} {\bibinfo  {journal} {Nature}\ }\textbf {\bibinfo {volume} {617}},\ \bibinfo {pages} {61} (\bibinfo {year} {2023})}\BibitemShut {NoStop}%
\bibitem [{\citenamefont {Xia}\ \emph {et~al.}(2018)\citenamefont {Xia}, \citenamefont {Bian},\ and\ \citenamefont {Kais}}]{xiaElectronicStructureCalculations2018}%
  \BibitemOpen
  \bibfield  {author} {\bibinfo {author} {\bibfnamefont {R.}~\bibnamefont {Xia}}, \bibinfo {author} {\bibfnamefont {T.}~\bibnamefont {Bian}}, \ and\ \bibinfo {author} {\bibfnamefont {S.}~\bibnamefont {Kais}},\ }\href {\doibase 10.1021/acs.jpcb.7b10371} {\bibfield  {journal} {\bibinfo  {journal} {The Journal of Physical Chemistry B}\ }\textbf {\bibinfo {volume} {122}},\ \bibinfo {pages} {3384} (\bibinfo {year} {2018})}\BibitemShut {NoStop}%
\bibitem [{\citenamefont {Genin}\ \emph {et~al.}(2019)\citenamefont {Genin}, \citenamefont {Ryabinkin},\ and\ \citenamefont {Izmaylov}}]{geninQuantumChemistryQuantum2019}%
  \BibitemOpen
  \bibfield  {author} {\bibinfo {author} {\bibfnamefont {S.~N.}\ \bibnamefont {Genin}}, \bibinfo {author} {\bibfnamefont {I.~G.}\ \bibnamefont {Ryabinkin}}, \ and\ \bibinfo {author} {\bibfnamefont {A.~F.}\ \bibnamefont {Izmaylov}},\ }\href {\doibase 10.48550/arXiv.1901.04715} {\enquote {\bibinfo {title} {Quantum chemistry on quantum annealers},}\ } (\bibinfo {year} {2019}),\ \Eprint {http://arxiv.org/abs/1901.04715} {arXiv:1901.04715 [physics, physics:quant-ph]} \BibitemShut {NoStop}%
\bibitem [{\citenamefont {Teplukhin}\ \emph {et~al.}(2020{\natexlab{a}})\citenamefont {Teplukhin}, \citenamefont {Kendrick}, \citenamefont {Tretiak},\ and\ \citenamefont {Dub}}]{teplukhinElectronicStructureDirect2020}%
  \BibitemOpen
  \bibfield  {author} {\bibinfo {author} {\bibfnamefont {A.}~\bibnamefont {Teplukhin}}, \bibinfo {author} {\bibfnamefont {B.~K.}\ \bibnamefont {Kendrick}}, \bibinfo {author} {\bibfnamefont {S.}~\bibnamefont {Tretiak}}, \ and\ \bibinfo {author} {\bibfnamefont {P.~A.}\ \bibnamefont {Dub}},\ }\href {\doibase 10.1038/s41598-020-77315-4} {\bibfield  {journal} {\bibinfo  {journal} {Scientific Reports}\ }\textbf {\bibinfo {volume} {10}},\ \bibinfo {pages} {20753} (\bibinfo {year} {2020}{\natexlab{a}})}\BibitemShut {NoStop}%
\bibitem [{\citenamefont {Teplukhin}\ \emph {et~al.}(2019)\citenamefont {Teplukhin}, \citenamefont {Kendrick},\ and\ \citenamefont {Babikov}}]{teplukhinCalculationMolecularVibrational2019}%
  \BibitemOpen
  \bibfield  {author} {\bibinfo {author} {\bibfnamefont {A.}~\bibnamefont {Teplukhin}}, \bibinfo {author} {\bibfnamefont {B.~K.}\ \bibnamefont {Kendrick}}, \ and\ \bibinfo {author} {\bibfnamefont {D.}~\bibnamefont {Babikov}},\ }\href {\doibase 10.1021/acs.jctc.9b00402} {\bibfield  {journal} {\bibinfo  {journal} {Journal of Chemical Theory and Computation}\ }\textbf {\bibinfo {volume} {15}},\ \bibinfo {pages} {4555} (\bibinfo {year} {2019})}\BibitemShut {NoStop}%
\bibitem [{\citenamefont {Teplukhin}\ \emph {et~al.}(2021)\citenamefont {Teplukhin}, \citenamefont {Kendrick}, \citenamefont {Mniszewski}, \citenamefont {Zhang}, \citenamefont {Kumar}, \citenamefont {Negre}, \citenamefont {Anisimov}, \citenamefont {Tretiak},\ and\ \citenamefont {Dub}}]{teplukhinComputingMolecularExcited2021}%
  \BibitemOpen
  \bibfield  {author} {\bibinfo {author} {\bibfnamefont {A.}~\bibnamefont {Teplukhin}}, \bibinfo {author} {\bibfnamefont {B.~K.}\ \bibnamefont {Kendrick}}, \bibinfo {author} {\bibfnamefont {S.~M.}\ \bibnamefont {Mniszewski}}, \bibinfo {author} {\bibfnamefont {Y.}~\bibnamefont {Zhang}}, \bibinfo {author} {\bibfnamefont {A.}~\bibnamefont {Kumar}}, \bibinfo {author} {\bibfnamefont {C.~F.~A.}\ \bibnamefont {Negre}}, \bibinfo {author} {\bibfnamefont {P.~M.}\ \bibnamefont {Anisimov}}, \bibinfo {author} {\bibfnamefont {S.}~\bibnamefont {Tretiak}}, \ and\ \bibinfo {author} {\bibfnamefont {P.~A.}\ \bibnamefont {Dub}},\ }\href {\doibase 10.1038/s41598-021-98331-y} {\bibfield  {journal} {\bibinfo  {journal} {Scientific Reports}\ }\textbf {\bibinfo {volume} {11}},\ \bibinfo {pages} {18796} (\bibinfo {year} {2021})}\BibitemShut {NoStop}%
\bibitem [{\citenamefont {Teplukhin}\ \emph {et~al.}(2020{\natexlab{b}})\citenamefont {Teplukhin}, \citenamefont {Kendrick},\ and\ \citenamefont {Babikov}}]{teplukhinSolvingComplexEigenvalue2020}%
  \BibitemOpen
  \bibfield  {author} {\bibinfo {author} {\bibfnamefont {A.}~\bibnamefont {Teplukhin}}, \bibinfo {author} {\bibfnamefont {B.~K.}\ \bibnamefont {Kendrick}}, \ and\ \bibinfo {author} {\bibfnamefont {D.}~\bibnamefont {Babikov}},\ }\href {\doibase 10.1039/D0CP04272B} {\bibfield  {journal} {\bibinfo  {journal} {Physical Chemistry Chemical Physics}\ }\textbf {\bibinfo {volume} {22}},\ \bibinfo {pages} {26136} (\bibinfo {year} {2020}{\natexlab{b}})}\BibitemShut {NoStop}%
\bibitem [{\citenamefont {Bravyi}\ \emph {et~al.}(2017)\citenamefont {Bravyi}, \citenamefont {Gambetta}, \citenamefont {Mezzacapo},\ and\ \citenamefont {Temme}}]{bravyiTaperingQubitsSimulate2017}%
  \BibitemOpen
  \bibfield  {author} {\bibinfo {author} {\bibfnamefont {S.}~\bibnamefont {Bravyi}}, \bibinfo {author} {\bibfnamefont {J.~M.}\ \bibnamefont {Gambetta}}, \bibinfo {author} {\bibfnamefont {A.}~\bibnamefont {Mezzacapo}}, \ and\ \bibinfo {author} {\bibfnamefont {K.}~\bibnamefont {Temme}},\ }\href {\doibase 10.48550/arXiv.1701.08213} {\enquote {\bibinfo {title} {Tapering off qubits to simulate fermionic {{Hamiltonians}}},}\ } (\bibinfo {year} {2017}),\ \Eprint {http://arxiv.org/abs/1701.08213} {arXiv:1701.08213 [quant-ph]} \BibitemShut {NoStop}%
\bibitem [{\citenamefont {Setia}\ \emph {et~al.}(2020)\citenamefont {Setia}, \citenamefont {Chen}, \citenamefont {Rice}, \citenamefont {Mezzacapo}, \citenamefont {Pistoia},\ and\ \citenamefont {Whitfield}}]{setiaReducingQubitRequirements2020}%
  \BibitemOpen
  \bibfield  {author} {\bibinfo {author} {\bibfnamefont {K.}~\bibnamefont {Setia}}, \bibinfo {author} {\bibfnamefont {R.}~\bibnamefont {Chen}}, \bibinfo {author} {\bibfnamefont {J.~E.}\ \bibnamefont {Rice}}, \bibinfo {author} {\bibfnamefont {A.}~\bibnamefont {Mezzacapo}}, \bibinfo {author} {\bibfnamefont {M.}~\bibnamefont {Pistoia}}, \ and\ \bibinfo {author} {\bibfnamefont {J.~D.}\ \bibnamefont {Whitfield}},\ }\href {\doibase 10.1021/acs.jctc.0c00113} {\bibfield  {journal} {\bibinfo  {journal} {Journal of Chemical Theory and Computation}\ }\textbf {\bibinfo {volume} {16}},\ \bibinfo {pages} {6091} (\bibinfo {year} {2020})}\BibitemShut {NoStop}%
\bibitem [{\citenamefont {Picozzi}\ and\ \citenamefont {Tennyson}(2023)}]{picozziSymmetryadaptedEncodingsQubit2023}%
  \BibitemOpen
  \bibfield  {author} {\bibinfo {author} {\bibfnamefont {D.}~\bibnamefont {Picozzi}}\ and\ \bibinfo {author} {\bibfnamefont {J.}~\bibnamefont {Tennyson}},\ }\href {\doibase 10.1088/2058-9565/acd86c} {\bibfield  {journal} {\bibinfo  {journal} {Quantum Science and Technology}\ }\textbf {\bibinfo {volume} {8}},\ \bibinfo {pages} {035026} (\bibinfo {year} {2023})}\BibitemShut {NoStop}%
\bibitem [{\citenamefont {Streif}\ \emph {et~al.}(2019)\citenamefont {Streif}, \citenamefont {Neukart},\ and\ \citenamefont {Leib}}]{streifSolvingQuantumChemistry2019}%
  \BibitemOpen
  \bibfield  {author} {\bibinfo {author} {\bibfnamefont {M.}~\bibnamefont {Streif}}, \bibinfo {author} {\bibfnamefont {F.}~\bibnamefont {Neukart}}, \ and\ \bibinfo {author} {\bibfnamefont {M.}~\bibnamefont {Leib}},\ }\href {\doibase 10.48550/arXiv.1811.05256} {\enquote {\bibinfo {title} {Solving {{Quantum Chemistry Problems}} with a {{D-Wave Quantum Annealer}}},}\ } (\bibinfo {year} {2019}),\ \Eprint {http://arxiv.org/abs/1811.05256} {arXiv:1811.05256 [quant-ph]} \BibitemShut {NoStop}%
\bibitem [{\citenamefont {Copenhaver}\ \emph {et~al.}(2021)\citenamefont {Copenhaver}, \citenamefont {Wasserman},\ and\ \citenamefont {{Wehefritz-Kaufmann}}}]{copenhaverUsingQuantumAnnealers2021}%
  \BibitemOpen
  \bibfield  {author} {\bibinfo {author} {\bibfnamefont {J.}~\bibnamefont {Copenhaver}}, \bibinfo {author} {\bibfnamefont {A.}~\bibnamefont {Wasserman}}, \ and\ \bibinfo {author} {\bibfnamefont {B.}~\bibnamefont {{Wehefritz-Kaufmann}}},\ }\href {\doibase 10.1063/5.0030397} {\bibfield  {journal} {\bibinfo  {journal} {The Journal of Chemical Physics}\ }\textbf {\bibinfo {volume} {154}},\ \bibinfo {pages} {034105} (\bibinfo {year} {2021})}\BibitemShut {NoStop}%
\bibitem [{\citenamefont {Seeley}\ \emph {et~al.}(2012)\citenamefont {Seeley}, \citenamefont {Richard},\ and\ \citenamefont {Love}}]{seeleyBravyiKitaevTransformationQuantum2012}%
  \BibitemOpen
  \bibfield  {author} {\bibinfo {author} {\bibfnamefont {J.~T.}\ \bibnamefont {Seeley}}, \bibinfo {author} {\bibfnamefont {M.~J.}\ \bibnamefont {Richard}}, \ and\ \bibinfo {author} {\bibfnamefont {P.~J.}\ \bibnamefont {Love}},\ }\href {\doibase 10.1063/1.4768229} {\bibfield  {journal} {\bibinfo  {journal} {The Journal of Chemical Physics}\ }\textbf {\bibinfo {volume} {137}},\ \bibinfo {pages} {224109} (\bibinfo {year} {2012})}\BibitemShut {NoStop}%
\bibitem [{\citenamefont {Tranter}\ \emph {et~al.}(2015)\citenamefont {Tranter}, \citenamefont {Sofia}, \citenamefont {Seeley}, \citenamefont {Kaicher}, \citenamefont {McClean}, \citenamefont {Babbush}, \citenamefont {Coveney}, \citenamefont {Mintert}, \citenamefont {Wilhelm},\ and\ \citenamefont {Love}}]{tranterBravyiKitaevTransformation2015}%
  \BibitemOpen
  \bibfield  {author} {\bibinfo {author} {\bibfnamefont {A.}~\bibnamefont {Tranter}}, \bibinfo {author} {\bibfnamefont {S.}~\bibnamefont {Sofia}}, \bibinfo {author} {\bibfnamefont {J.}~\bibnamefont {Seeley}}, \bibinfo {author} {\bibfnamefont {M.}~\bibnamefont {Kaicher}}, \bibinfo {author} {\bibfnamefont {J.}~\bibnamefont {McClean}}, \bibinfo {author} {\bibfnamefont {R.}~\bibnamefont {Babbush}}, \bibinfo {author} {\bibfnamefont {P.~V.}\ \bibnamefont {Coveney}}, \bibinfo {author} {\bibfnamefont {F.}~\bibnamefont {Mintert}}, \bibinfo {author} {\bibfnamefont {F.}~\bibnamefont {Wilhelm}}, \ and\ \bibinfo {author} {\bibfnamefont {P.~J.}\ \bibnamefont {Love}},\ }\href {\doibase 10.1002/qua.24969} {\bibfield  {journal} {\bibinfo  {journal} {International Journal of Quantum Chemistry}\ }\textbf {\bibinfo {volume} {115}},\ \bibinfo {pages} {1431} (\bibinfo {year} {2015})}\BibitemShut {NoStop}%
\bibitem [{\citenamefont {Bravyi}\ and\ \citenamefont {Kitaev}(2002)}]{bravyiFermionicQuantumComputation2002}%
  \BibitemOpen
  \bibfield  {author} {\bibinfo {author} {\bibfnamefont {S.~B.}\ \bibnamefont {Bravyi}}\ and\ \bibinfo {author} {\bibfnamefont {A.~Y.}\ \bibnamefont {Kitaev}},\ }\href {\doibase 10.1006/aphy.2002.6254} {\bibfield  {journal} {\bibinfo  {journal} {Annals of Physics}\ }\textbf {\bibinfo {volume} {298}},\ \bibinfo {pages} {210} (\bibinfo {year} {2002})}\BibitemShut {NoStop}%
\bibitem [{\citenamefont {Sun}\ \emph {et~al.}(2018)\citenamefont {Sun}, \citenamefont {Berkelbach}, \citenamefont {Blunt}, \citenamefont {Booth}, \citenamefont {Guo}, \citenamefont {Li}, \citenamefont {Liu}, \citenamefont {McClain}, \citenamefont {Sayfutyarova}, \citenamefont {Sharma}, \citenamefont {Wouters},\ and\ \citenamefont {Chan}}]{sunPySCFPythonbasedSimulations2018}%
  \BibitemOpen
  \bibfield  {author} {\bibinfo {author} {\bibfnamefont {Q.}~\bibnamefont {Sun}}, \bibinfo {author} {\bibfnamefont {T.~C.}\ \bibnamefont {Berkelbach}}, \bibinfo {author} {\bibfnamefont {N.~S.}\ \bibnamefont {Blunt}}, \bibinfo {author} {\bibfnamefont {G.~H.}\ \bibnamefont {Booth}}, \bibinfo {author} {\bibfnamefont {S.}~\bibnamefont {Guo}}, \bibinfo {author} {\bibfnamefont {Z.}~\bibnamefont {Li}}, \bibinfo {author} {\bibfnamefont {J.}~\bibnamefont {Liu}}, \bibinfo {author} {\bibfnamefont {J.~D.}\ \bibnamefont {McClain}}, \bibinfo {author} {\bibfnamefont {E.~R.}\ \bibnamefont {Sayfutyarova}}, \bibinfo {author} {\bibfnamefont {S.}~\bibnamefont {Sharma}}, \bibinfo {author} {\bibfnamefont {S.}~\bibnamefont {Wouters}}, \ and\ \bibinfo {author} {\bibfnamefont {G.~K.-L.}\ \bibnamefont {Chan}},\ }\href {\doibase 10.1002/wcms.1340} {\bibfield  {journal} {\bibinfo  {journal} {WIREs Computational Molecular Science}\ }\textbf {\bibinfo {volume} {8}},\ \bibinfo {pages} {e1340} (\bibinfo {year} {2018})}\BibitemShut {NoStop}%
\bibitem [{\citenamefont {McClean}\ \emph {et~al.}(2020)\citenamefont {McClean}, \citenamefont {Rubin}, \citenamefont {Sung}, \citenamefont {Kivlichan}, \citenamefont {{Bonet-Monroig}}, \citenamefont {Cao}, \citenamefont {Dai}, \citenamefont {Fried}, \citenamefont {Gidney}, \citenamefont {Gimby}, \citenamefont {Gokhale}, \citenamefont {H{\"a}ner}, \citenamefont {Hardikar}, \citenamefont {Havl{\'i}{\v c}ek}, \citenamefont {Higgott}, \citenamefont {Huang}, \citenamefont {Izaac}, \citenamefont {Jiang}, \citenamefont {Liu}, \citenamefont {McArdle}, \citenamefont {Neeley}, \citenamefont {O'Brien}, \citenamefont {O'Gorman}, \citenamefont {Ozfidan}, \citenamefont {Radin}, \citenamefont {Romero}, \citenamefont {Sawaya}, \citenamefont {Senjean}, \citenamefont {Setia}, \citenamefont {Sim}, \citenamefont {Steiger}, \citenamefont {Steudtner}, \citenamefont {Sun}, \citenamefont {Sun}, \citenamefont {Wang}, \citenamefont {Zhang},\ and\ \citenamefont {Babbush}}]{mccleanOpenFermionElectronicStructure2020}%
  \BibitemOpen
  \bibfield  {author} {\bibinfo {author} {\bibfnamefont {J.~R.}\ \bibnamefont {McClean}}, \bibinfo {author} {\bibfnamefont {N.~C.}\ \bibnamefont {Rubin}}, \bibinfo {author} {\bibfnamefont {K.~J.}\ \bibnamefont {Sung}}, \bibinfo {author} {\bibfnamefont {I.~D.}\ \bibnamefont {Kivlichan}}, \bibinfo {author} {\bibfnamefont {X.}~\bibnamefont {{Bonet-Monroig}}}, \bibinfo {author} {\bibfnamefont {Y.}~\bibnamefont {Cao}}, \bibinfo {author} {\bibfnamefont {C.}~\bibnamefont {Dai}}, \bibinfo {author} {\bibfnamefont {E.~S.}\ \bibnamefont {Fried}}, \bibinfo {author} {\bibfnamefont {C.}~\bibnamefont {Gidney}}, \bibinfo {author} {\bibfnamefont {B.}~\bibnamefont {Gimby}}, \bibinfo {author} {\bibfnamefont {P.}~\bibnamefont {Gokhale}}, \bibinfo {author} {\bibfnamefont {T.}~\bibnamefont {H{\"a}ner}}, \bibinfo {author} {\bibfnamefont {T.}~\bibnamefont {Hardikar}}, \bibinfo {author} {\bibfnamefont {V.}~\bibnamefont {Havl{\'i}{\v c}ek}}, \bibinfo {author} {\bibfnamefont {O.}~\bibnamefont {Higgott}}, \bibinfo {author}
  {\bibfnamefont {C.}~\bibnamefont {Huang}}, \bibinfo {author} {\bibfnamefont {J.}~\bibnamefont {Izaac}}, \bibinfo {author} {\bibfnamefont {Z.}~\bibnamefont {Jiang}}, \bibinfo {author} {\bibfnamefont {X.}~\bibnamefont {Liu}}, \bibinfo {author} {\bibfnamefont {S.}~\bibnamefont {McArdle}}, \bibinfo {author} {\bibfnamefont {M.}~\bibnamefont {Neeley}}, \bibinfo {author} {\bibfnamefont {T.}~\bibnamefont {O'Brien}}, \bibinfo {author} {\bibfnamefont {B.}~\bibnamefont {O'Gorman}}, \bibinfo {author} {\bibfnamefont {I.}~\bibnamefont {Ozfidan}}, \bibinfo {author} {\bibfnamefont {M.~D.}\ \bibnamefont {Radin}}, \bibinfo {author} {\bibfnamefont {J.}~\bibnamefont {Romero}}, \bibinfo {author} {\bibfnamefont {N.~P.~D.}\ \bibnamefont {Sawaya}}, \bibinfo {author} {\bibfnamefont {B.}~\bibnamefont {Senjean}}, \bibinfo {author} {\bibfnamefont {K.}~\bibnamefont {Setia}}, \bibinfo {author} {\bibfnamefont {S.}~\bibnamefont {Sim}}, \bibinfo {author} {\bibfnamefont {D.~S.}\ \bibnamefont {Steiger}}, \bibinfo {author} {\bibfnamefont
  {M.}~\bibnamefont {Steudtner}}, \bibinfo {author} {\bibfnamefont {Q.}~\bibnamefont {Sun}}, \bibinfo {author} {\bibfnamefont {W.}~\bibnamefont {Sun}}, \bibinfo {author} {\bibfnamefont {D.}~\bibnamefont {Wang}}, \bibinfo {author} {\bibfnamefont {F.}~\bibnamefont {Zhang}}, \ and\ \bibinfo {author} {\bibfnamefont {R.}~\bibnamefont {Babbush}},\ }\href {\doibase 10.1088/2058-9565/ab8ebc} {\bibfield  {journal} {\bibinfo  {journal} {Quantum Science and Technology}\ }\textbf {\bibinfo {volume} {5}},\ \bibinfo {pages} {034014} (\bibinfo {year} {2020})}\BibitemShut {NoStop}%
\bibitem [{\citenamefont {Hehre}\ \emph {et~al.}(1969)\citenamefont {Hehre}, \citenamefont {Stewart},\ and\ \citenamefont {Pople}}]{hehreSelfConsistentMolecularOrbitalMethods1969}%
  \BibitemOpen
  \bibfield  {author} {\bibinfo {author} {\bibfnamefont {W.~J.}\ \bibnamefont {Hehre}}, \bibinfo {author} {\bibfnamefont {R.~F.}\ \bibnamefont {Stewart}}, \ and\ \bibinfo {author} {\bibfnamefont {J.~A.}\ \bibnamefont {Pople}},\ }\href {\doibase 10.1063/1.1672392} {\bibfield  {journal} {\bibinfo  {journal} {The Journal of Chemical Physics}\ }\textbf {\bibinfo {volume} {51}},\ \bibinfo {pages} {2657} (\bibinfo {year} {1969})}\BibitemShut {NoStop}%
\bibitem [{\citenamefont {Binkley}\ \emph {et~al.}(1980)\citenamefont {Binkley}, \citenamefont {Pople},\ and\ \citenamefont {Hehre}}]{binkleySelfconsistentMolecularOrbital1980}%
  \BibitemOpen
  \bibfield  {author} {\bibinfo {author} {\bibfnamefont {J.~S.}\ \bibnamefont {Binkley}}, \bibinfo {author} {\bibfnamefont {J.~A.}\ \bibnamefont {Pople}}, \ and\ \bibinfo {author} {\bibfnamefont {W.~J.}\ \bibnamefont {Hehre}},\ }\href {\doibase 10.1021/ja00523a008} {\bibfield  {journal} {\bibinfo  {journal} {Journal of the American Chemical Society}\ }\textbf {\bibinfo {volume} {102}},\ \bibinfo {pages} {939} (\bibinfo {year} {1980})}\BibitemShut {NoStop}%
\bibitem [{\citenamefont {Ditchfield}\ \emph {et~al.}(1971)\citenamefont {Ditchfield}, \citenamefont {Hehre},\ and\ \citenamefont {Pople}}]{ditchfieldSelfConsistentMolecularOrbitalMethods1971}%
  \BibitemOpen
  \bibfield  {author} {\bibinfo {author} {\bibfnamefont {R.}~\bibnamefont {Ditchfield}}, \bibinfo {author} {\bibfnamefont {W.~J.}\ \bibnamefont {Hehre}}, \ and\ \bibinfo {author} {\bibfnamefont {J.~A.}\ \bibnamefont {Pople}},\ }\href {\doibase 10.1063/1.1674902} {\bibfield  {journal} {\bibinfo  {journal} {The Journal of Chemical Physics}\ }\textbf {\bibinfo {volume} {54}},\ \bibinfo {pages} {724} (\bibinfo {year} {1971})}\BibitemShut {NoStop}%
\bibitem [{\citenamefont {Dunning}(1989)}]{dunningGaussianBasisSets1989}%
  \BibitemOpen
  \bibfield  {author} {\bibinfo {author} {\bibfnamefont {T.~H.}\ \bibnamefont {Dunning}, \bibfnamefont {Jr.}},\ }\href {\doibase 10.1063/1.456153} {\bibfield  {journal} {\bibinfo  {journal} {The Journal of Chemical Physics}\ }\textbf {\bibinfo {volume} {90}},\ \bibinfo {pages} {1007} (\bibinfo {year} {1989})}\BibitemShut {NoStop}%
\end{thebibliography}%


\begin{thebibliography}{1}

\bibitem{copenhaverUsingQuantumAnnealers2021}
Justin Copenhaver, Adam Wasserman, and Birgit {Wehefritz-Kaufmann}.
\newblock Using quantum annealers to calculate ground state properties of molecules.
\newblock {\em The Journal of Chemical Physics}, 154(3):034105, January 2021.

\bibitem{mccleanOpenFermionElectronicStructure2020}
Jarrod~R. McClean, Nicholas~C. Rubin, Kevin~J. Sung, Ian~D. Kivlichan, Xavier {Bonet-Monroig}, Yudong Cao, Chengyu Dai, E.~Schuyler Fried, Craig Gidney, Brendan Gimby, Pranav Gokhale, Thomas H{\"a}ner, Tarini Hardikar, Vojt{\v e}ch Havl{\'i}{\v c}ek, Oscar Higgott, Cupjin Huang, Josh Izaac, Zhang Jiang, Xinle Liu, Sam McArdle, Matthew Neeley, Thomas O'Brien, Bryan O'Gorman, Isil Ozfidan, Maxwell~D. Radin, Jhonathan Romero, Nicolas P.~D. Sawaya, Bruno Senjean, Kanav Setia, Sukin Sim, Damian~S. Steiger, Mark Steudtner, Qiming Sun, Wei Sun, Daochen Wang, Fang Zhang, and Ryan Babbush.
\newblock {{OpenFermion}}: The electronic structure package for quantum computers.
\newblock {\em Quantum Science and Technology}, 5(3):034014, June 2020.

\bibitem{streifSolvingQuantumChemistry2019}
Michael Streif, Florian Neukart, and Martin Leib.
\newblock Solving {{Quantum Chemistry Problems}} with a {{D-Wave Quantum Annealer}}, March 2019.

\bibitem{xiaElectronicStructureCalculations2018}
Rongxin Xia, Teng Bian, and Sabre Kais.
\newblock Electronic {{Structure Calculations}} and the {{Ising Hamiltonian}}.
\newblock {\em The Journal of Physical Chemistry B}, 122(13):3384--3395, April 2018.

\end{thebibliography}

\end{document}


\maketitle

\subsection*{Equivalence of Our Implementation of the Xia-Bian-Kais Method}

Here, we demonstrate that our implementation of the Xia-Bian-Kais (XBK) method produces results analogous to those of \cite{copenhaverUsingQuantumAnnealers2021}, \cite{xiaElectronicStructureCalculations2018}, and \cite{streifSolvingQuantumChemistry2019}. Since the original authors of the XBK method do not provide the method they use to calculate their potential energy surfaces, we assume that they use a classical method \cite{xiaElectronicStructureCalculations2018}. In \cite{copenhaverUsingQuantumAnnealers2021}, the authors develop an implementation of the XBK method that both applies the qubit-to-Ising transformation and iterates over the possible sign permutations to find the minimal energy. They do this in tandem with either a simulated annealer or a quantum annealer with interfacing the Ocean SDK Python package developed by D-Wave Systems. Here, we use the simulated annealing sampler, as demonstrated in \cite{copenhaverUsingQuantumAnnealers2021} there is only a vanishing difference between the simulated and quantum annealer for the small molecules considered. 

\begin{figure*}[h!]
    \centering
    \includegraphics[width=\textwidth]{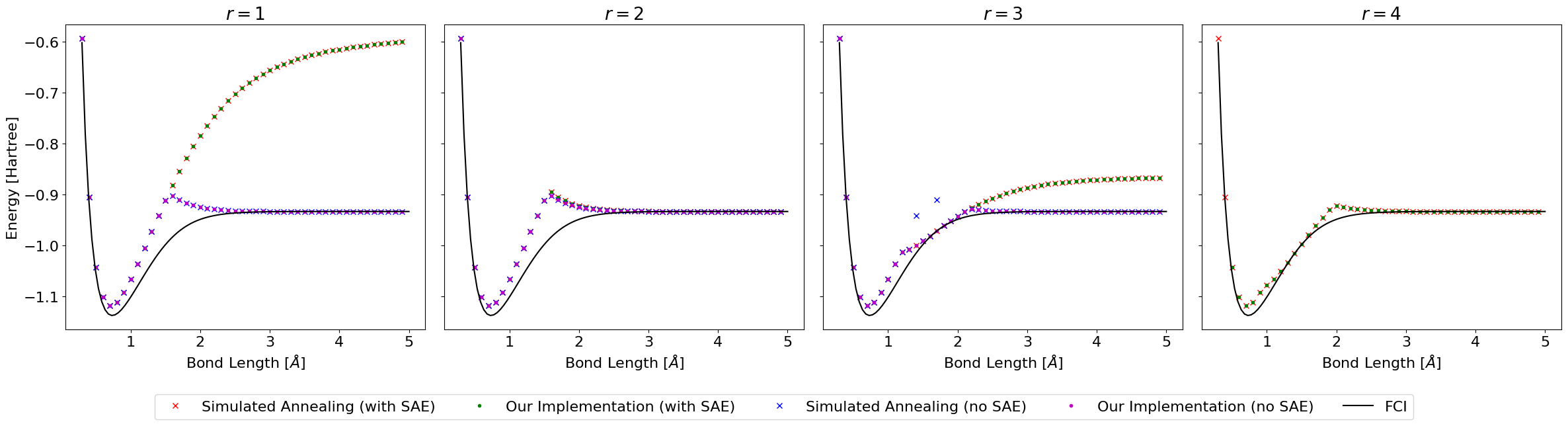}
    \caption{The potential energy surface of H$_2$ in STO-3G, computed with both our implementation and the implementation of \cite{copenhaverUsingQuantumAnnealers2021} of the XBK method. }
    \label{sfig_h2}
\end{figure*}

\begin{figure*}[h!]
    \centering
    \includegraphics[width=\textwidth]{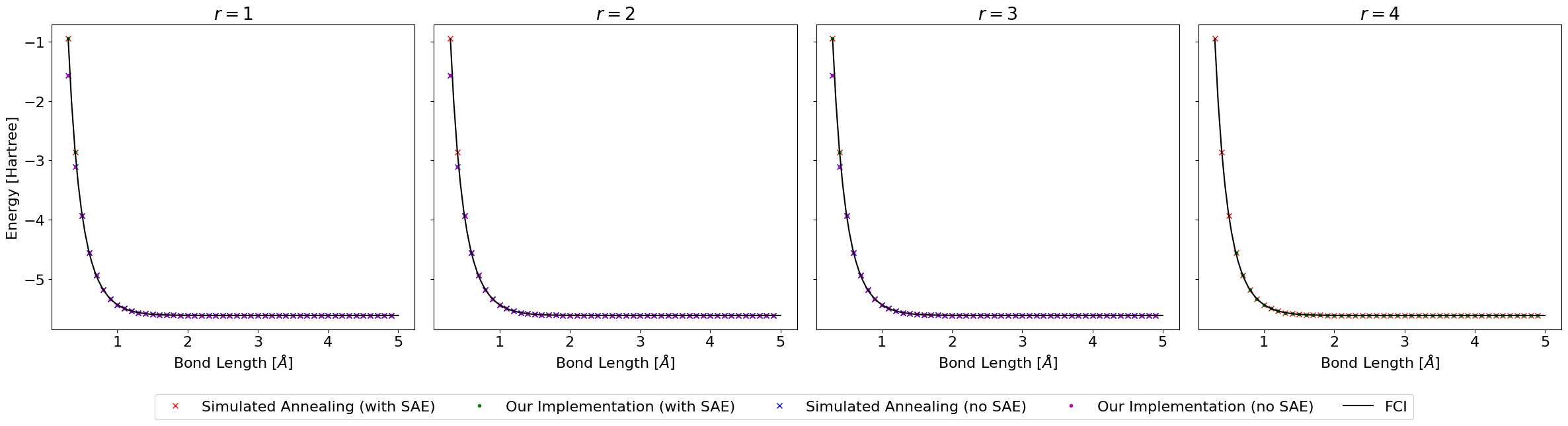}
    \caption{The potential energy surface of He$_2$ in STO-3G, computed with both our implementation and the implementation of \cite{copenhaverUsingQuantumAnnealers2021} of the XBK method.}
    \label{sfig_he2}
\end{figure*}

In Fig. \ref{sfig_h2} and Fig. \ref{sfig_he2}, we demonstrate that our implementation of the XBK method reproduces the same values as the method used in \cite{copenhaverUsingQuantumAnnealers2021}. In Fig. \ref{sfig_h2}, we show that our method matches exactly the potential energy curves created for H$_2$ for $r\in\{1,2,3,4\}$ with and without symmetry-adapted encodings. In Fig. \ref{sfig_he2}, we show this for He$_2$ for the same $r$ values considered. We attempted calculations for the other molecules considered in our work, but the implementation in \cite{copenhaverUsingQuantumAnnealers2021} was not able to properly converge for at least a 12-qubit problem (the non-symmetry-adapted LiH Ising Hamiltonian in STO-3G) and thus was unable to calculate the potential energy surface of any other of the molecules considered.

We do note that at some points, our implementation performs better than the method developed in \cite{copenhaverUsingQuantumAnnealers2021}. This is because while the method in \cite{copenhaverUsingQuantumAnnealers2021} uses an annealing routine, our method uses numerically exact diagonalization. Therefore, our implementation is able to find the ground state exactly without any probabilistic errors, while the implementation in \cite{copenhaverUsingQuantumAnnealers2021} uses a probabilistic heuristic and thus occasionally suffers from (sometimes large) random errors. 

Therefore, for all the cases considered, our implementation of the XBK method matches the implementation created in \cite{copenhaverUsingQuantumAnnealers2021} or performs better. The better performance can be explained in terms of the non-deterministic behavior of the simulated and quantum annealing algorithms compared to the numerically exact nature of exact diagonalization. Furthermore, this serves as evidence that our algorithm better serves as a pure investigation of the XBK method, as compared to incorporating probabilistic and hardware errors associated with simulated and quantum annealing with previous work \cite{streifSolvingQuantumChemistry2019}$^,$\cite{copenhaverUsingQuantumAnnealers2021}.

\subsection*{Effect of Basis Sets on the Potential Energy Surface of H$_2$}

We present the results of our method applied to molecular hydrogen (H$_2$) in other basis sets here. We consider the STO-3G and STO-6G basis sets, as well as the 3-21G, 6-31G, and cc-pVDZ basis sets here. From Table S1, we see that when we use cc-pVDZ, we require 15 qubits to represent the $r=1$ symmetry-adapted Ising Hamiltonian. For the 3-21G and 6-31G basis sets, we require 5 qubits. Therefore, we are limited to considering small $r$ for both of these basis sets.

\begin{figure*}[h!]
    \centering
    \includegraphics[width=0.6\textwidth]{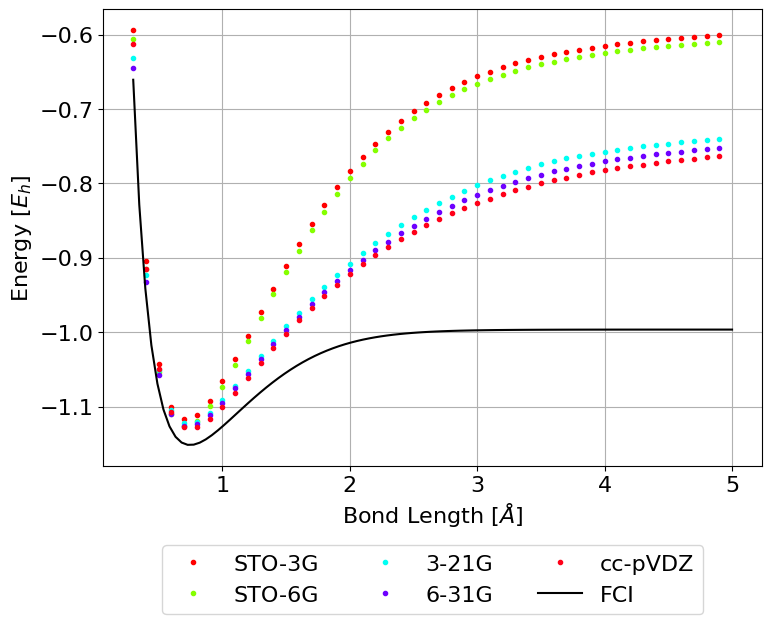}
    \caption{The potential energy surface of the symmetry-adapted Ising Hamiltonian of H$_2$ at $r=1$ in different basis sets.}
    \label{sfigh2basisr1}
\end{figure*}

\begin{figure*}[h!]
    \centering
    \includegraphics[width=0.6\textwidth]{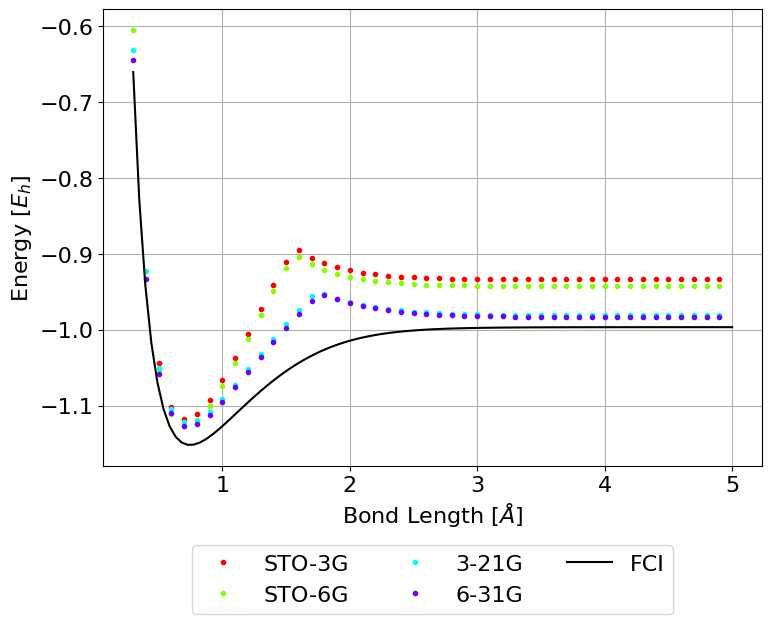}
    \caption{The potential energy surface of the symmetry-adapted Ising Hamiltonian of H$_2$ at $r=2$ in different basis sets.}
    \label{sfigh2basisr2}
\end{figure*}

In Fig. \ref{sfigh2basisr1} and Fig. \ref{sfigh2basisr2}, we see that while the larger basis sets do perform better than the minimal STO-3G basis set, the improvement may not be worth the increased number of qubits used to construct the Hamiltonians. Comparing Fig. \ref{sfigh2basisr1} with Fig. 2 of the main text, we see that we are able to construct the potential energy curve of H$_2$ with the $r=16$, STO-3G symmetry-adapted Hamiltonian (containing 16 qubits) more accurately than with the $r=1$, cc-pVDZ symmetry-adapted Hamiltonian (containing 15 qubits). Because the use of larger basis sets drastically increases the number of qubits needed to construct the qubit Hamiltonians, therefore limiting the maximal $R$ value one can use, the decision to use a larger basis set is not necessarily a major improvement. Although the use of larger basis sets does increase the accuracy of the potential energy surface, there are more significant increases in accuracy from directly increasing $r$.

\subsection*{Potential Energy Surfaces for Additional Molecules}

With current classical computational constraints, our implementation can easily handle systems with up to 16 qubits, i.e., $2^{16} \times 2^{16}$ dimensional Hermitian matrices. Because we use symmetry-adapted encodings (SAE) that can reduce up to 5 qubits, this means that the largest molecules that we can consider with current hardware have at most a 21-qubit Hamiltonian that maintains a $\mathbb{Z}_2^5$ symmetry. At these numbers of qubits, we are only able to perform the symmetry-adapted calculation and thus are not able to compare them to the non-symmetry-adapted case. Furthermore, this prevents us from commenting on the full eigenspectrum of the Ising Hamiltonians before and after the application of SAE. Hence, we do not include these cases in the main text and instead present them here. 

In addition to H$_2$, He$_2$, LiH, H$_2$O, BH$_3$, and NH$_3$, we also calculate the potential energy surfaces of N$_2$, O$_2$, CO, F$_2$, Li$_2$, and CH$_4$. We find it critical to note that for some of the systems considered, the FCI calculation we perform in PySCF does not converge at many bond lengths considered. Therefore, discontinuities may exist with the FCI curve - these are not phyiscal discontinuities and instead arise solely from the method used in PySCF.

\begin{figure*}[h!]
    \centering
    \begin{subfigure}[t]{0.40\textwidth}
        \centering
        \includegraphics[width=\textwidth]{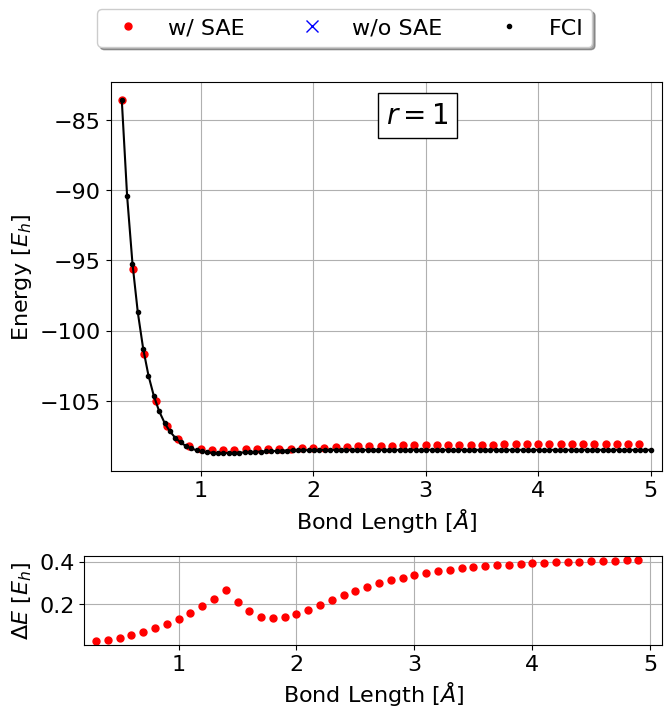}
        \caption{The PES of N$_2$ in STO-6G at $r=1$.}
        \label{fig:si3}
    \end{subfigure}
    \hfill
    \begin{subfigure}[t]{0.40\textwidth}
        \centering
        \includegraphics[width=\textwidth]{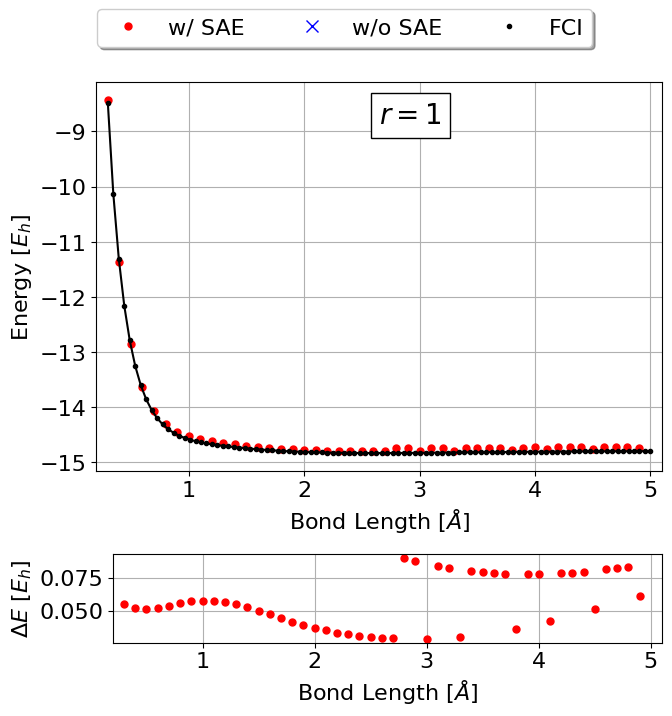}
        \caption{The PES of Li$_2$ in STO-6G at $r=1$.}
        \label{fig:si4}
    \end{subfigure}

    \vspace{0.5cm}

    \begin{subfigure}[t]{0.40\textwidth}
        \centering
        \includegraphics[width=\textwidth]{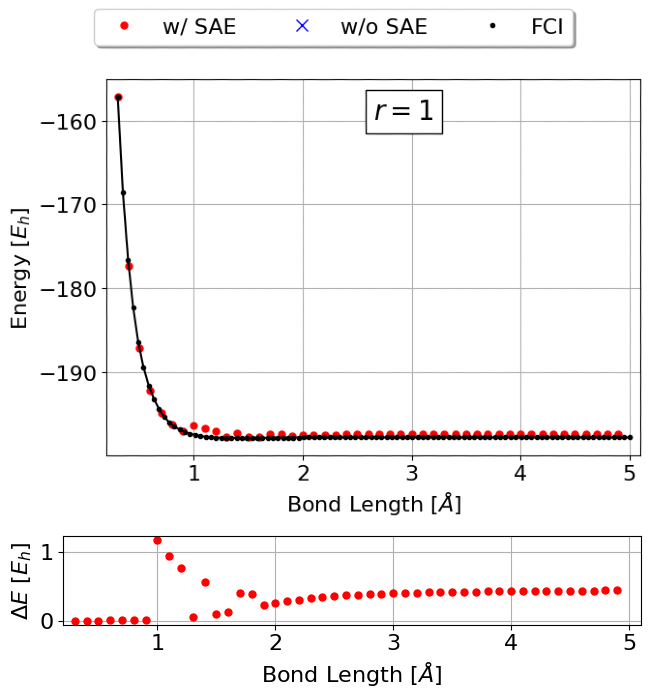}
        \caption{The PES of F$_2$ in STO-6G at $r=1$.}
        \label{fig:si5}
    \end{subfigure}
    \hfill
    \begin{subfigure}[t]{0.40\textwidth}
        \centering
        \includegraphics[width=\textwidth]{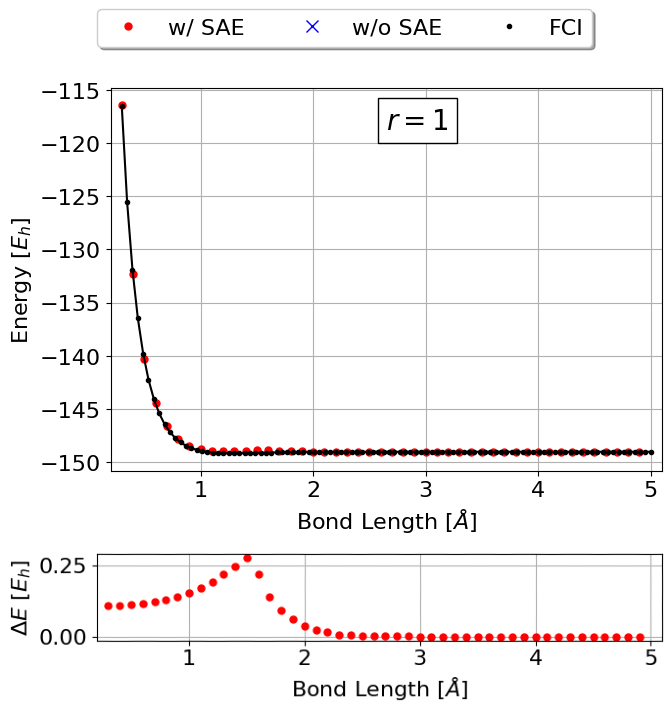}
        \caption{The PES of O$_2$ in STO-6G at $r=1$.}
        \label{fig:si6}
    \end{subfigure}

    \vspace{0.5cm}

    \begin{subfigure}[t]{0.40\textwidth}
        \centering
        \includegraphics[width=\textwidth]{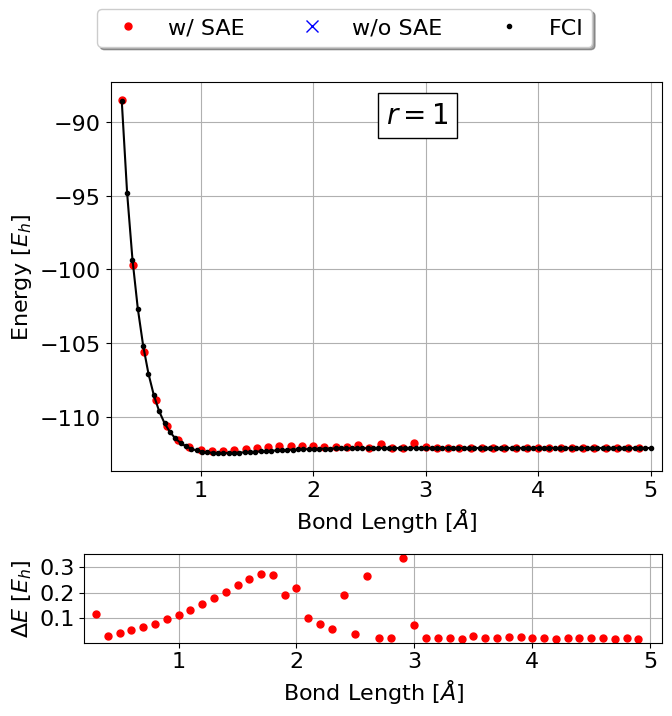}
        \caption{The PES of CO in STO-6G at $r=1$.}
        \label{fig:si7}
    \end{subfigure}
    \hfill
    \begin{subfigure}[t]{0.40\textwidth}
        \centering
        \includegraphics[width=\textwidth]{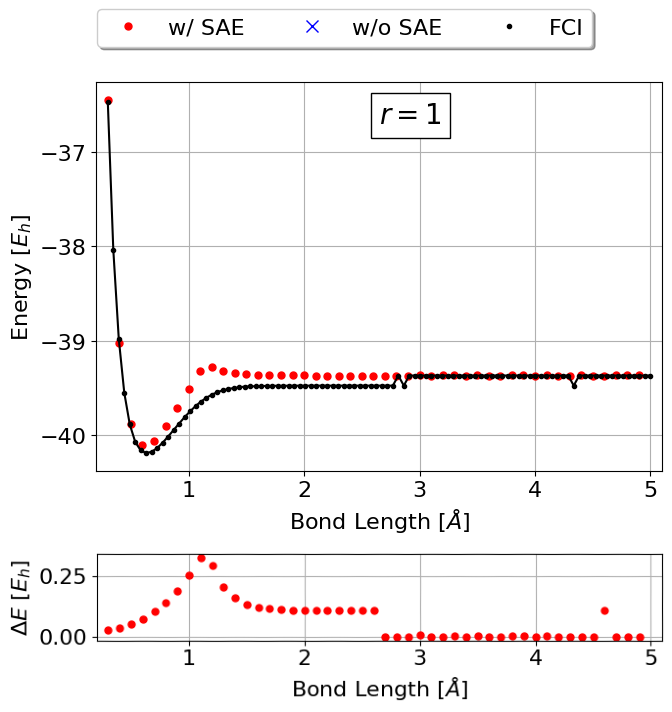}
        \caption{The PES of CH$_4$ in STO-6G at $r=1$.}
        \label{fig:si8}
    \end{subfigure}
    
    \caption{The potential energy surfaces (PES) of molecules not considered in the main text. }
    \label{fig:grid}
\end{figure*}

We show our results for these molecules in Fig. \ref{fig:grid}. Based on the findings of the main text, we expect that symmetry-adapted Ising Hamiltonian ground state energies of N$_2$, F$_2$, and Li$_2$ should differ significantly from the non-symmetry-adapted case because of the number of electrons each atomic species has in its dissociation limit. This is corroborated by the  behavior we see in Fig. \ref{fig:grid}, as for these three molecules we have non-converging behavior at the large bond-length limit, indicating that the necessary states for this limit are missing. 

We also expect that for CH$_4$, CO, and O$_2$ that symmetry-adapted encodings should work well, as these molecules dissociate into atomic speices that do not all contain an odd number of electrons. We see this explicitly, as for each case the differnece between the symmetry-adapted Ising Hamiltonian ground state energy vanishes in the large-bond length limit, converging proplerly to the FCI energy. Therefore, for these molecules, we are able to effectively model the transition that occurs as one extends the bond-length from the molecular phase to the atomic phase.  

Because we are unable to compute the non-symmetry-adapted Ising Hamiltonian ground state energies for these molecules (see Table S1) as they contain too many qubits for current hardware, we are unable to investigate the differences in the higher-order eigenspectrum for the two cases. Therefore, we are unable to directly claim that the molecules that we expect to have emerging errors between symmetry-adapted Ising ground states and non-symmetry-adapted Ising ground states actually do suffer these errors. We, therefore, are only able to discuss the converging behavior as seen above. However, on the basis of this convergence behavior we see similar trends to the molecules presented in the main text, and therefore, we see these trends as further evidence of our claims. 

\subsection*{Qubit Counts}

In Table S1, we present the minimum number of qubits needed to simulate H$_2$, LiH, He$_2$, H$_2$O, O$_2$, N$_2$, Li$_2$, F$_2$, CO, BH$_3$ NH$_3$, and CH$_4$ in the STO-3G, 3-21G, and cc-pVDZ basis sets. We note that for our molecules considered, we can directly extend this to include the STO-6G and 6-31G basis sets. This is because for both the STO-3G and STO-6G basis sets, the same number of qubits are required to construct the qubit Hamiltonian. This pattern extends to 3-21G and 6-31G. Therefore, to identify how many qubits are required to simulate a particular molecule in the STO-6G or the 6-31G basis sets, simply look at the corresponding STO-3G or 3-21G entry in Table S1.

\begin{table}[ht]
\centering
\renewcommand{\arraystretch}{1.3}
\begin{tabular}{|c|c|c|c|c|}
\hline

\textbf{Molecule} & \textbf{Bool. Sym.} & \textbf{Qubits [STO-3G]} & \textbf{Qubits [3-21G]} & \textbf{Qubits [cc-pVDZ]}\\ \hline
H$_2$ & $\mathbb{Z}_2^3$ & 1 & 5 & 15\\ 
H$_2$O & $\mathbb{Z}_2^4$ & 10 & 22 & 44 \\ 
LiH & $\mathbb{Z}_2^4$ & 8 & 18 & 34 \\ 
He$_2$ & $\mathbb{Z}_2^3$ & 1 & 5 & 15\\ 
O$_2$ & $\mathbb{Z}_2^5$ & 15 & 31 & 51 \\ 
N$_2$ & $\mathbb{Z}_2^5$ & 15 & 31 & 51\\ 
Li$_2$ & $\mathbb{Z}_2^5$ & 15 & 31 & 51\\ 
BH$_3$ & $\mathbb{Z}_2^4$ & 12 & 26 & 54\\ 
NH$_3$ & $\mathbb{Z}_2^3$ & 13 & 27 & 55\\ 
CH$_4$ & $\mathbb{Z}_2^4$ & 14 & 30 & 64\\ 
CO & $\mathbb{Z}_2^4$ & 16 & 32 & 52\\ 
F$_2$ & $\mathbb{Z}_2^5$ & 15 & 31 & 52\\ \hline
\end{tabular}
\caption{Qubit requirements for various molecules using the STO-3G and 3-21G basis sets with a bond length of 1.0 Å.}
\label{tab:qubit_requirements}
\end{table}

The number of qubits needed to simulate each molecule in each basis set was calculated by using OpenFermion's count\_qubits() function from their utils module\cite{mccleanOpenFermionElectronicStructure2020}. For each molecule, we construct the qubit Hamiltonian using the JW encoding based on the full Boolean symmetry group. We also find that even though the Hamiltonians for the same molecule in a larger basis set require more qubits, the number of qubits saved via symmetry-adapted encodings is a constant, independent of basis set size. Therefore, in order to find the number of qubits needed to construct the non-symmetry-adapted Hamiltonians, simply take the order of the Boolean symmetry group and add it directly to the number of qubits required to construct the Hamiltonian in a particular basis set. 

\clearpage
\bibliographystyle{plain}
\bibliography{references}